%% file: jmcbtd_arxiv.tex
\documentclass[10pt,a4paper,twocolumn,hidelinks]{article} 
\usepackage[ a4paper, top=20mm, bottom=22mm, left=18mm, right=18mm, columnsep=8mm, footskip=10mm ]{geometry}
\usepackage{microtype}
\usepackage{balance} 
\usepackage{booktabs}
\usepackage{array}
\usepackage[caption=false,font=normalsize,labelfont=sf,textfont=sf]{subfig}
\usepackage{textcomp}
\usepackage{stfloats}
\usepackage{url}
\usepackage{verbatim}
\usepackage{graphicx}
\usepackage[short]{optidef}
\usepackage{hyperref}
\input{preamble.tex}

\usepackage{graphicx} 
\usepackage{booktabs} 
\usepackage{caption} 
\usepackage{titlesec} 
\titleformat{\section} {\normalfont\bfseries} {\thesection.} {0.55em} {} 
\titleformat{\subsection} {\normalfont} {\thesubsection.} {0.55em} {} 
\titleformat{\subsubsection} {\normalfont\itshape} {\thesubsubsection.} {0.55em} {} 
\titlespacing*{\section} {0pt} {2.4ex plus 0.5ex minus 0.2ex} {0.9ex plus 0.2ex} 
\titlespacing*{\subsection} {0pt} {1.8ex plus 0.4ex minus 0.2ex} {0.7ex plus 0.2ex} 
\usepackage[bottom]{footmisc} 
\usepackage{titling} 
\pretitle{ \begin{center} \fontsize{15}{18}\selectfont \bfseries }
\posttitle{ \par \end{center} \vspace{0.5em} }
\preauthor{ \begin{center} \normalsize }
\postauthor{ \par \end{center} \vspace{-3.5em} }

\newenvironment{paperabstract} {%
\begin{quote} \small \noindent\textbf{Abstract -- }\ignorespaces } {%
    \end{quote} }

\newenvironment{paperkeywords} {%
\begin{quote} \small \noindent\textbf{Keywords -- }\ignorespaces } {%
    \end{quote} }

\newcommand{\blfootnote}[1]{%
\begingroup \renewcommand{\thefootnote}{}%
\footnotetext{#1}%
\endgroup }

\title{\bfseries Joint Higher-Order Low-Rank Tensor Decompositions for Multichannel Acoustic System
    Compression } 
\author{
    Matthias Blochberger$^{1\star}$,
    Filip Elvander$^2$,
    Jan Østergaard$^3$,
    Toon van Waterschoot$^1$ \\[0.5em] 
    \normalsize $^1$STADIUS, Department of Electrical Engineering, KU Leuven, Leuven, Belgium \\
    \normalsize $^2$Department of Information and Communications Engineering, Aalto
    University, Espoo, Finland \\
    \normalsize $^3$Department of Electronic Systems, Aalborg University, Aalborg, Denmark 
}
\date{} 
\begin{document} 
\twocolumn[ 
    \maketitle 
    \begin{paperabstract} 
            \input{jmcbtd_abstract.tex}
    \end{paperabstract}
    \begin{paperkeywords}
        \input{jmcbtd_keywords.tex}
    \end{paperkeywords}
] 
\blfootnote{$^*$Corresponding author: matthias.blochberger@kuleuven.be}
\blfootnote{This manuscript is currently under peer review.}

\input{jmcbtd_manuscript.tex}

\section*{Funding}
    This research work was carried out at the ESAT Laboratory of KU Leuven, and has
    received funding from KU Leuven internal funds (C14/21/075), Research Foundation Flanders
    (FWO projects S005525N and G0A0424N), the Flemish Government (AI Research Program), and
    the European Union's Horizon 2020 research and innovation programme under the Marie
    Skłodowska-Curie grant agreement No. 956369.  
    The scientific responsibility is assumed by the authors.
\balance 
\bibliographystyle{IEEEtran}
\bibliography{ieeectl,dist-low-rank-est.bib}

\end{document}

%% file: preamble.tex
\usepackage{tcolorbox}
\tcbuselibrary{theorems}
\newtcbtheorem{mynote}{Note}
{colback=blue!5,colframe=blue!35!black,fonttitle=\bfseries}{th}
\newtcbtheorem{mytodo}{Todo}
{colback=red!5,colframe=red!60!black,fonttitle=\bfseries}{th}

\usepackage{amsmath,amssymb,amsfonts,bm}
\usepackage{nicematrix}
\usepackage{algorithm}
\usepackage{algpseudocode}
\usepackage{cases}
\usepackage{tabularx}
\usepackage{multirow}
\usepackage{tikz}
\usepackage[only,llbracket,rrbracket]{stmaryrd}

\newcommand{\appref}[1]{\hyperref[#1]{Appendix~\ref*{#1}}}

\usepackage[acronym]{glossaries}
\loadglsentries[type]{acro.tex}

\newcommand*{\mtxb}[1]{\bm{\mathrm{#1}}}
\newcommand*{\vect}[1]{\bm{\mathrm{#1}}}
\newcommand*{\matr}[1]{\bm{\mathrm{#1}}}
\newcommand*{\tens}[1]{\bm{\mathcal{#1}}}
\newcommand*{\set}[1]{\mathcal{#1}}
\newcommand*{\nprod}[1]{\times_{#1}}
\newcommand*{\T}{{\top}}

\newcommand*{\naturals}{\mathbb{N}}
\newcommand*{\reals}{\mathbb{R}}
\newcommand*{\complex}{\mathbb{C}}
\newcommand*{\Ree}[1]{\operatorname{Re} \bigl[ #1 \bigr]}

\newcommand{\mybigcirc}{\mathop{\raisebox{-0.4ex}{\scalebox{1.8}{$\circ$}}}\displaylimits}

\newcommand*{\oprod}[2]{#1 \circ #2} 

\newcommand*{\lsv}[2]{\operatorname{LSV}\!\!_{#1}\bigl(#2\bigr)} 

\newcommand*{\h}{\mtxb{h}}

\def\mybigtimes{\mathop{\mathchoice{
   \vcenter{\hbox to10bp{\vrule height15bp width0pt \pdfliteral{
   q 1 J .8 w 0 1 m 10 14 l S 0 14 m 10 1 l S Q
}\hss}}}{
   \vcenter{\hbox to10bp{\kern1bp\vrule height10bp width0pt \pdfliteral{
   q 1 J .65 w 0 0 m 8 10 l S 0 10 m 8 0 l S Q
}\hss}}}{\times}{\times}
}}

\definecolor{tab-blue}{HTML}{1f77b4}
\definecolor{tab-orange}{HTML}{ff7f0e}
\definecolor{tab-green}{HTML}{2ca02c}
\definecolor{tab-red}{HTML}{d62728}
\definecolor{tab-purple}{HTML}{9467bd}
\definecolor{tab-brown}{HTML}{8c564b}
\definecolor{tab-pink}{HTML}{e377c2}
\definecolor{tab-gray}{HTML}{7f7f7f}
\definecolor{tab-olive}{HTML}{bcbd22}
\definecolor{tab-cyan}{HTML}{17becf}

%% file: acro.tex
\newacronym{wgn}{WGN}{white Gaussian noise}
\newacronym{ces}{CES}{complex elliptically symmetric}
\newacronym{scm}{SCM}{sample covariance matrix}
\newacronym{dof}{DoF}{degrees of freedom}

\newacronym{dft}{DFT}{discrete Fourier transform}
\newacronym{dtft}{DTFT}{discrete time Fourier transform}
\newacronym{fft}{FFT}{fast Fourier transform}

\newacronym{rir}{RIR}{room impulse response}
\newacronym{iir}{IIR}{infinite impulse response}
\newacronym{fir}{FIR}{finite impulse response}
\newacronym{ism}{ISM}{image source method}
\newacronym{rism}{RISM}{randomized image source method}
\newacronym{ar}{AR}{autoregressive}

\newacronym{edc}{EDC}{energy decay curve}

\newacronym{smard}{SMARD}{single- and multichannel audio recordings database}
\newacronym{sriracha}{SRIRACHA}{shoebox room impulse response archive with varying absorption}
\newacronym{churchir}{ChurchIR}{church impulse response dataset}

\newacronym{szc}{SZC}{sound zone control}
\newacronym{vast}{VAST}{variable span trade-off filter}
\newacronym{acc}{ACC}{acoustic contrast control}
\newacronym{pm}{PM}{pressure matching}
\newacronym{aec}{AEC}{acoustic echo cancellation}
\newacronym{anc}{ANC}{active noise control}

\newacronym{qos}{QoS}{quality of service}
\newacronym{fpi}{FPI}{fixed point iteration}

\newacronym{mse}{MSE}{mean square error}
\newacronym{nmse}{NMSE}{normalized mean square error}
\newacronym{nm}{NM}{normalized misalignment}
\newacronym{npm}{NPM}{normalized projection misalignment}
\newacronym{ac}{AC}{acoustic contrast}
\newacronym{sd}{SD}{signal distortion}
\newacronym{stoi}{STOI}{short term objective intelligibility}
\newacronym{re}{RE}{residual energy}
\newacronym{sinr}{SINR}{signal-to-interference-plus-noise ratio}
\newacronym{snr}{SNR}{signal-to-noise ratio}
\newacronym{tee}{TEE}{tail envelope error}
\newacronym{ler}{LER}{log error ratio}

\newacronym{evd}{EVD}{eigenvalue decomposition}
\newacronym{gevd}{GEVD}{generalized eigenvalue decomposition}
\newacronym{pevd}{PEVD}{polynomial eigenvalue decomposition}
\newacronym{pgevd}{PGEVD}{polynomial generalized eigenvalue decomposition}
\newacronym{qr}{QR}{QR decomposition}
\newacronym{svd}{SVD}{singular value decomposition}
\newacronym{hosvd}{HOSVD}{higher-order singular value decomposition}
\newacronym{lmlra}{LMLRA}{low multilinear rank approximation}
\newacronym{hooi}{HOOI}{higher-order orthogonal iteration}
\newacronym{mwf}{MWF}{multichannel Wiener filter}
\newacronym{atc}{ACT}{adapt-then-combine}
\newacronym{cta}{CTA}{combine-then-adapt}
\newacronym{cg}{CG}{conjugate gradient}

\newacronym{cp}{CPD}{canonical polyadic decomposition}
\newacronym{ccp}{C-CPD}{coupled canonical polyadic decomposition}
\newacronym{tuck}{TD}{Tucker decomposition}
\newacronym{ctuck}{C-TD}{coupled Tucker decomposition}
\newacronym{btd}{BTD}{block term decomposition}
\newacronym{cbtd}{C-BTD}{coupled block term decomposition}
\newacronym{sbtd}{SMCBTD}{stacked multichannel block term decomposition}
\newacronym{jmcbtd}{JMCBTD}{joint multichannel block term decomposition}
\newacronym{glram}{GLRAM}{generalized low-rank approximation of matrices}
\newacronym{kpgnobtd}{SeCO‑BTD}{Separable-Curvature Online BTD}

\newacronym{pz}{PZ}{pole-zero}

\newacronym{ola}{OLA}{overlap-add}
\newacronym{ols}{OLS}{overlap-save}
\newacronym{wola}{WOLA}{weighted overlap-add}

\newacronym{ls}{LS}{least-squares}
\newacronym{lm}{LM}{Levenberg-Marquardt}
\newacronym{gn}{GN}{Gauss-Newton}
\newacronym{lms}{LMS}{least mean squares}
\newacronym{nlms}{NLMS}{normalized least mean squares}
\newacronym{fxlms}{FxLMS}{filtered-x least mean squares}
\newacronym{rls}{RLS}{recursive least squares}
\newacronym{rlstot}{RLS-TOT}{recursive least squares for third-order tensor decomposition}
\newacronym{als}{ALS}{alternating least squares}
\newacronym{rals}{RALS}{regularized alternating least squares}
\newacronym{mdf}{MDF}{multidelay block frequency-domain adaptive filter}
\newacronym{pbfdaf}{PBFDAF}{partitioned-block frequency-domain adaptive filter}

\newacronym{pseq}{PSEQ}{perfect sequence}
\newacronym{ops}{OPS}{orthogonal periodic sequence}
\newacronym{pps}{PPS}{perfect periodic sequence}

\newacronym{mnf}{MNF}{minimum noise fraction}
\newacronym{rkhs}{RKHS}{reproducing kernel Hilbert space}
\newacronym{krr}{KRR}{kernel ridge regression}

\newacronym{psd}{PSD}{positive semi-definite}
\newacronym{sdr}{SDR}{semi-definite relaxation}
\newacronym{sdp}{SDP}{semi-definite program}
\newacronym{qcqp}{QCQP}{quadratically constrained quadratic program}

\newacronym{mttkrp}{MTTKRP}{matricized-tensor times Khatri-Rao product}
\newacronym{flop}{FLOP}{floating point operation}

\newacronym{wsn}{WSN}{wireless sensor network}
\newacronym{wasn}{WASN}{wireless acoustic sensor network}
\newacronym{iot}{IoT}{Internet of Things}
\newacronym{admm}{ADMM}{alternating direction method of multipliers}
\newacronym{bsi}{BSI}{blind system identification}
\newacronym{si}{SI}{system identification}
\newacronym{rtf}{RTF}{relative transfer function}
\newacronym{simo}{SIMO}{single-input multiple-output}
\newacronym{cap}{CAP}{common acoustical pole}

%% file: jmcbtd_abstract.tex
    Multichannel \glspl{rir} in acoustics are often long and therefore
    the sample count becomes large if seen as \gls{fir} filters, making storage and
    real-time convolution costly.  
    \Gls{pz} models are an established way to represent \glspl{rir} compactly as
    \gls{iir} transfer functions.  
    We show that a \gls{pz} model implies a low-rank \gls{cp} of tensorized \glspl{rir}:
    the early reflections form a sparse rank-1 sum, while the late reverberation admits a
    \gls{cp} with factors determined by the poles. This motivates
    higher-order-tensor-based compression of \glspl{rir} via the \gls{btd}, a
    generalization of the \gls{cp}.  
    We propose two multichannel compression schemes: (i) a \gls{sbtd} applied to a
    (D+1)-mode tensor of \glspl{rir}, and (ii) a \gls{jmcbtd} that shares subsets of
    factors across channels discovered by spectral clustering on factor–subspace affinity
    graphs.  
    Both are optimized by an alternating \gls{hooi} procedure.  
    On simulated rooms as well as measured datasets, the proposed methods achieve
    consistently lower reconstruction error at equal compression ratios compared to
    separate single-channel decompositions and a \gls{glram} baseline.  
    The largest gains occur with scalar cores (i.e., \gls{cp}), supporting the
    \gls{pz}-to-\gls{cp} modelling link.  
    We also find D = 3 to be a practical sweet spot,
    offering most of the achievable gain at moderate compute/memory cost. The compressed
    models enable efficient downstream multichannel convolution.

%% file: jmcbtd_keywords.tex
    Room impulse response, tensor decomposition, block-term decomposition, canonical
    polyadic decomposition, spectral clustering

%% file: jmcbtd_manuscript.tex
\glsresetall

\section{Introduction}\label{sec:joint_comp:introduction}  
In 
room acoustic signal processing, the \gls{rir} is a fundamental
concept that describes the point-to-point acoustic path from a sound source to a
microphone within a room \cite{kuttruffRoomAcoustics2009}.  
It captures the effects of the room's geometry, materials, and potentially interior
furnishings on the sound propagation, including reflections, diffractions, and
absorptions.  
They are of great importance in various applications aiming to measure, evaluate, model or
compensate for the acoustic properties of a room, such as in source localization
\cite{eversLOCATAChallengeAcoustic2020}, speech enhancement and source separation
\cite{gannotConsolidatedPerspectiveMultimicrophone2017}, 
room geometry estimation \cite{macwilliamSimultaneousAcousticEcho2023}, or dereverberation
\cite{naylorSpeechDereverberation2010}.  

The \glspl{rir} are typically modelled as either \gls{iir} filters (e.g.,
\cite{mourjopoulosPoleZeroModeling1991,vairettiScalableAlgorithmPhysically2017}) or
\gls{fir} filters, depending on the application and the desired properties of the model.  
The \gls{fir} model is often preferred due to its simplicity, with the drawback
of requiring a large number of parameters.    
For an average-sized room, the \gls{rir} can be several thousands of samples, while for
large spaces, such as concert halls or churches, a \gls{rir} can be several tens to
hundreds of thousands of samples long.  
This can lead to issues with storage and computational complexity, especially when
multiple \glspl{rir} are considered.  
This is particularly relevant in applications that aim to reproduce the room in a virtual
environment, such as virtual reality or augmented reality, where a potentially large
number of \glspl{rir} measured within the acoustic space are needed.  
While \gls{iir} models are compact and can represent long \glspl{rir} with few parameters,
they tend to be more difficult to work with, e.g., due to numerical stability issues.  
The \gls{pz} model \cite{mourjopoulosPoleZeroModeling1991} is a common
approach to model \glspl{rir} as \gls{iir} filters.  
It enables a mathematical representation of the room modes, the resonances of the room,
through its poles.  
The room modes are shared by all \glspl{rir} in a room
\cite{hanedaCommonAcousticalPole1994}.  
The method proposed in \cite{vanwaterschootDistributedEstimationEqualization2012} takes
advantage of common poles in a distributed estimation algorithm.  
Other literature uses the common-pole model for spatial interpolation of \glspl{rir}
\cite{dasRoomImpulseResponse2021} by estimating modal parameters.  
In this paper, we show that there is a clear connection between the \gls{pz} model and
low-rank tensor approximations by means of modal factorization of \glspl{rir}.  

The need for compact representation of \glspl{rir} has led to the development of
various compression techniques, with recent ones focusing on low-rank decompositions of
\glspl{rir}.  
The truncated \gls{svd} is a well-known method for low-rank approximation of matrices,
which has been applied to matricized \glspl{rir} 
\cite{huangDimensionalityReductionRoom2023a}.  
In \cite{jalmbyMultiChannelLowRankConvolution2024}, a generalized form of the \gls{svd},
the \gls{glram} \cite{yeGeneralizedLowRank2004} is used to jointly compress a set of
matricized \glspl{rir}.  
We use the term "matricized" to describe the arrangement of consecutive \gls{rir} segments
of appropriate length as columns of an appropriately sized matrix; the analogue term we
use for higher-order tensors is "tensorized".  
Higher-order decompositions, such as the Kronecker product decomposition or the \gls{cp}
have been applied to individual \glspl{rir} arranged in tensor form
\cite{jalmbyCompressionRoomImpulse2024, huangAcousticSystemIdentification2022}.  
All these methods focus on compact representations that allow efficient downstream uses
of the compressed \glspl{rir}, mainly described in
\cite{jalmbyCompressionRoomImpulse2024,jalmbyMultiChannelLowRankConvolution2024} where the
decompositions yield fast convolution algorithms.  
Other compression approaches for multichannel acoustic systems include
\cite{mezzaLargeScaleRoomImpulse2024} where neural audio codecs are applied.  
Although this approach is effective in terms of compression, the particular application
does not consider joint compression of multiple \glspl{rir} nor does the compact
representation yield more efficient downstream applications.  
While an effective approach, it does not consider the same requirements and is not
considered in our comparisons.  

We extend these low-rank compression approaches to a joint multichannel setting by
introducing tensor-decomposition-based models for sets of \glspl{rir}. 
Our framework is based on the \acrfull{btd}, which provides a convenient umbrella model
for several standard low-rank tensor decompositions. 
In particular, the \gls{tuck}, also known as \gls{lmlra}, is the natural extension of
matrix low-rank approximations such as \gls{glram} 
\cite{jalmbyMultiChannelLowRankConvolution2024} to arbitrary \(D\)-mode tensors. 
The \gls{cp} can be viewed as a restricted \gls{tuck} model with a superdiagonal core
whose dimensions are all equal to the \gls{cp} rank
\cite{koldaTensorDecompositionsApplications2009}. 
The \gls{btd} further generalizes these models by representing a tensor as a sum of
block terms, where each block is itself a \gls{tuck} model with its own core and factor
matrices.
This hierarchy allows us to treat \gls{cp}, \gls{tuck}, and intermediate \gls{btd} models
within a common framework.  
It is particularly useful in the present context because the
\gls{pz}-based analysis suggests a low-rank tensor structure for tensorized \glspl{rir},
but does not determine a priori whether scalar-core, single-core, or multi-block
structures are most parameter-efficient in practice.  
By comparing \gls{cp}, \gls{tuck}, and \gls{btd} models experimentally, we assess which
member of this decomposition family best matches measured and simulated \glspl{rir}.  
As shown in Section~\ref{sec:joint_comp:numerical-simulations}, scalar-core \gls{btd},
i.e., \gls{cp}, provides the most parameter-efficient representation in many of the
considered scenarios, supporting the proposed connection between the \gls{pz} model and
low-rank \gls{cp} structure.


The main contribution of this paper is the introduction of two joint compression methods.  
First, we propose multichannel compression by applying \gls{btd} to a \((D\!+\!1)\)-mode
tensor, i.e., a stacked tensor of \(D\)-mode tensorized \glspl{rir}, which we refer to as
\gls{sbtd}.  
Then we introduce a coupled \gls{btd}, named \gls{jmcbtd}, for which we formulate the
problem as a joint multichannel tensor decomposition problem, where subsets of
decomposition factors are shared across \glspl{rir}.  
The set of shared factors across subsets of channels is determined by a spectral
clustering approach.  
The factors of individual \glspl{rir} are first computed using a standard \gls{btd} and
then used to form graphs of decomposition factors within each mode and rank.  
The graphs have affinity edges between factors that are similar, i.e., have a small
subspace distance.  
Spectral clustering then forms clusters, i.e. subgraphs, of factors that can be considered
similar within a mode and rank.  
This differs from the \gls{ccp} approach analyzed in
\cite{sorensenCoupledCanonicalPolyadic2015a,sorensenCoupledCanonicalPolyadic2015} which
considers all rank factors of a mode to be shared across the decompositions of multiple
tensors, while we consider an arbitrary subset of factors across modes and ranks to be
shared, determined by spectral clustering.  
The optimization algorithm for \gls{btd} is based on alternating \glspl{hooi} 
\cite{
    koldaTensorDecompositionsApplications2009,
    delathauwerDecompositionsHigherOrderTensor2008b,
    delathauwerDecompositionsHigherOrderTensor2008a,
    delathauwerDecompositionsHigherOrderTensor2008}.

The paper is structured as follows:
In \autoref{sec:joint_comp:acoustic-room-impulse-reponses}, we introduce the basic concepts of
tensorization of \glspl{rir} and the \gls{btd}.  
In \autoref{sec:joint_comp:pole-zero-model-and-low-rank-tensor-representation}, the connection
between the \gls{pz} model of \glspl{rir} and the low-rank tensor representation is
established.  
In \autoref{sec:joint_comp:joint-multichannel-room-impulse-response-compression-using-btd}, we
propose the joint multichannel compression methods \gls{jmcbtd} and \gls{sbtd}.   
In \autoref{sec:joint_comp:numerical-simulations}, we evaluate the proposed methods on simulated and
measured data, showing that they outperform the state-of-the-art.  
In \autoref{sec:joint_comp:discussion}, we elaborate on a few points of discussion and future work,
and we conclude in \autoref{sec:joint_comp:conclusion}.  

\begin{figure}[h]
    \begin{center}
    \setlength{\fboxsep}{8pt}
    \noindent\fbox{%
        \parbox{0.92\columnwidth}{%
            \linespread{1.0}
            \footnotesize
            \hypertarget{notation}{\textbf{Notation:}}
We use lowercase letters for scalar values (\(x\)), lowercase bold letters for
vectors (\(\vect{x} \in \reals^{I}\)), uppercase bold letters for matrices (\(\matr{X}
\in \reals^{I_1 \times I_2}\)), and calligraphic bold letters for \(D\)-mode
tensors (\(\tens{X} \in \reals^{I_1 \times \cdots \times I_D}\)).  
Element-wise (multi-)indexing uses square brackets, e.g. \(\vect{x}[i]\), \(\matr{X}[i_1, i_2]\), and
\(\tens{X}[i_1,...,i_D]\) for the \(i\)-th, the \((i_1,i_2)\)-th, and the
\((i_1,...,i_D)\)-th element of the vector, matrix, and tensor, respectively.   
We use \(\oprod{}{}\) for the outer product of vectors, 
with multiple denoted as
\begin{equation*}
    \vect{x}_1 \circ \vect{x}_2 \circ \cdots \circ
    \vect{x}_D \equiv \mybigcirc_{d=1}^{D} \vect{x}_d = \tens{X} \in \reals^{I_1 \times \cdots \times I_D}
\end{equation*}
when \(\vect{x}_d \in \reals^{I_d}\) for \(d \in [D] = \{1, 2, \ldots, D\}\).  
Mode-\(d\) unfolding \(\matr{X}_{(d)} \in \reals^{I_d \times
\prod_{\delta \neq d} I_\delta}\) of \(\tens{X} \in \reals^{I_1 \times \cdots \times
I_D}\) stacks the mode-\(d\) fibres as columns.  
The \(d\)-mode product of a tensor \(\tens{X} \in \reals^{I_1 \times
\cdots \times I_D}\) with a matrix \(\matr{A} \in \reals^{J \times I_d}\) is a
contraction along mode \(d\) such that 
\begin{align*}
    \tens{Y} &= \tens{X} \nprod{d} \matr{A}
        \in \reals^{
            I_1 \times \cdots \times I_{d-1} \times J \times I_{d+1} \times \cdots \times I_D
        }\\
    \iff \matr{Y}_{(d)} &= \matr{A} \matr{X}_{(d)}
        \in \reals^{J \times \prod_{\delta \neq d} I_\delta}.
\end{align*}
We denote the multiple-mode product on a tensor with \(D \geq E\) modes as 
\begin{equation*}
    \tens{Y} = \tens{X} \nprod{1} \matr{A}_1 \cdots \nprod{E} \matr{A}_E \equiv
    \tens{X} \nprod{d \in [E]} \matr{A}_d.
\end{equation*}
We denote the multiple-mode product with all modes except \(d\) as
\begin{equation*}
    \tens{Y} = \tens{X} \nprod{\delta \in [D] \setminus \{d\}} \matr{A}_\delta \equiv
    \tens{X} \nprod{\delta \neq d} \matr{A}_\delta.
\end{equation*}
More details on these tensor specific definitions can be found in
\cite{koldaTensorDecompositionsApplications2009}.  

Further we will make use of the following short-hand:
For \(\matr{A} \in \reals^{I \times J}\) and \(1 \leq K \leq \min\{I, J\}\), define
\(\lsv{K}{\matr{A}} \in \reals^{I \times K}\) to be the matrix whose columns
are \(K\) leading left singular vectors of \(\matr{A}\).  
}%
}
\end{center}
\end{figure}

\section{Tensorized Acoustic Room Impulse Responses and Tensor Decompositions}
\label{sec:joint_comp:acoustic-room-impulse-reponses}
Let \(L\) be the number of samples of an \gls{rir} \(h_m(n)\) that are considered
significant above the system noise floor, and assume this length to be the same for
all \(m\), since the \glspl{rir} share the same reverberation \cite{kuttruffRoomAcoustics2009}.  

Tensor decompositions have been applied to \glspl{rir} in prior work:  
Considering matrices as \(2\)-mode tensors, \cite{huangDimensionalityReductionRoom2023a}
uses the truncated \gls{svd} to compress multichannel \glspl{rir} arranged as a matrix.  
\cite{jalmbyMultiChannelLowRankConvolution2024} uses \gls{glram} to jointly compress
multiple matricized \glspl{rir}.  
Higher-order tensor decompositions have been applied to individual \glspl{rir} in
\cite{jalmbyLowRankRoomImpulse2023,jalmbyCompressionRoomImpulse2024} where they are
arranged in \(D\)-mode tensors  and subsequently \gls{cp} is applied, as well as 
\cite{huangAcousticSystemIdentification2022} in which the authors use the Kronecker
decomposition, i.e., \gls{cp}, assuming that there are some decomposition factors that
stay constant across channels while others vary.  
The last example is similar in spirit to our proposed \gls{jmcbtd}.

\subsection{Tensorization}
\label{sec:joint_comp:tensorsization}
Choose positive integers
\(I_{1},I_{2},\dots,I_{D}\) such that
\begin{equation}
    L = \prod_{d=1}^{D} I_d.
\end{equation}
Define the tensorization bijection between \(n = 0,1,\dots,L-1\) and the multi‐index
\((i_{1},i_{2},\dots,i_{D})\) by
\begin{equation}
    \label{eq:multi-index}
    n  =  i_{1} + i_{2} I_{1} +  i_{3} \left( I_{1} I_{2} \right) 
        + \cdots + i_{D} \left( I_{1} I_{2} \cdots I_{D-1} \right),
\end{equation}
where \(0 \le i_{d} \le I_{d}-1\) for each \(d\).  
Then define the \(D\)-mode tensorized \gls{rir}
\(\tens{H}_m \in \reals^{I_{1} \times I_{2} \times \cdots \times I_{D}}\) by
\begin{equation}
    \begin{aligned}
        \tens{H}_m&[ i_{1}, i_{2}, \dots, i_{D} ]\\
        =& \,h_m \left(
                i_{1} + i_{2} I_{1} + i_{3} I_{1} I_{2} + \cdots+  i_{D} I_{1} I_{2} \cdots I_{D-1}
            \right).
    \end{aligned}\label{eq:rir-tensorization}
\end{equation}

\subsection{Block Term Decomposition}
\label{sec:joint_comp:btd}

Let \(\tens{H} \in \reals^{I_1 \times I_2 \times \cdots \times I_D}\) denote a tensorized
\gls{rir}.  
A \gls{btd} with \(R\) blocks approximates \(\tens{H}\) as 
\cite{koldaTensorDecompositionsApplications2009,delathauwerDecompositionsHigherOrderTensor2008b,delathauwerDecompositionsHigherOrderTensor2008a,delathauwerDecompositionsHigherOrderTensor2008}
\begin{equation}
    \tens{H} \approx \widehat{\tens{H}} = \sum_{r=1}^{R} \tens{G}^{(r)} \nprod{d \in [D]} \matr{A}^{(r,d)},
    \label{eq:joint_comp:btd}
\end{equation}
where \(\tens{G}^{(r)} \in \reals^{K_1^{(r)} \times \cdots \times K_D^{(r)}}\) is the core
tensor of block \(r\), and \(\matr{A}^{(r,d)} \in \reals^{I_d \times K_d^{(r)}}\) is its
factor matrix in mode \(d\).  
We consider the orthogonally constrained case
with \(K_d^{(r)} \leq I_d\).  
For compact notation, we define the set of block-mode indices as
\begin{equation}
    \set{I} \triangleq \{1,\ldots,R\} \times \{1,\ldots,D\}.
    \label{eq:block_mode_indices}
\end{equation}
The corresponding approximation problem is
\begin{mini}
    {\tens{G}^{(r)}, \matr{A}^{(r, d)}}{\Bigl\|
        \tens{H} - \sum_{r=1}^{R} \tens{G}^{(r)} \nprod{d \in [D]} \matr{A}^{(r, d)}
        \Bigr\|_{F}^2}
    {\label{eq:btd-minimization}}{}
    \addConstraint{
        \matr{A}^{(r, d) \T} \matr{A}^{(r, d)}
    }{= \matr{I}}{\quad \text{for all } (r,d) \in \set{I}},
\end{mini}

The \gls{btd} generalizes several standard tensor decompositions.  
If each core is scalar, \(\tens{G}^{(r)} = g^{(r)} \in \reals\), then
\eqref{eq:joint_comp:btd} reduces to the rank-\(R\) \gls{cp},
\begin{equation}
    \widehat{\tens{H}} = \sum_{r=1}^{R} g^{(r)} \vect{a}^{(r,1)} \circ \cdots \circ \vect{a}^{(r,D)}.
    \label{eq:cpd_special_case}
\end{equation}
If \(R=1\), then \eqref{eq:joint_comp:btd} reduces to the Tucker decomposition,
\begin{equation}
    \widehat{\tens{H}} = \tens{G} \nprod{d\in[D]} \matr{A}^{(d)}.
    \label{eq:td_special_case}
\end{equation}
The factors and cores in \eqref{eq:joint_comp:btd} can be computed using standard
alternating \gls{hooi}-type algorithms for \gls{btd}; we refer the reader to 
\cite{koldaTensorDecompositionsApplications2009,delathauwerDecompositionsHigherOrderTensor2008b,delathauwerDecompositionsHigherOrderTensor2008a,delathauwerDecompositionsHigherOrderTensor2008}
for details.  
In \autoref{sec:joint_comp:joint-multichannel-room-impulse-response-compression-using-btd},
we adapt this principle to the proposed multichannel decompositions with shared factors.

\section{Pole-Zero Model and Low-Rank Tensor Representation}
\label{sec:joint_comp:pole-zero-model-and-low-rank-tensor-representation}

To establish the connection between \gls{pz} models of \glspl{rir} and low-rank tensor
representations, consider the causal rational model
\begin{equation}
    y(n) = H(q)x(n) = \frac{B(q)}{A(q)}x(n),
\end{equation}
where \(H(q)\) represents the \gls{rir} \(h(n)\), \(B(q)\) and \(A(q)\) are numerator and
denominator polynomials of degrees \(P_B\) and \(P_A\), respectively, and \(q^{-l}x(n) =
x(n-l)\). In the \(z\)-domain,
\begin{equation}
    H(z)
    =
    \frac{B(z)}{A(z)}
    =
    \frac{b_0 + b_1z^{-1} + \cdots + b_{P_B}z^{-P_B}}
    {1 + a_1z^{-1} + \cdots + a_{P_A}z^{-P_A}},
\end{equation}
with real coefficients \(a_k,b_k\).  
We assume, for simplicity, that the poles \(z_r\), \(r=1,\ldots,P_A\), are simple. Since
\(h(n) \in \reals\), complex poles occur in conjugate pairs.

\subsection{Modal Factorization of Impulse Responses}
\label{sec:joint_comp:modal-factorization}
The decomposition used below is the discrete-time counterpart of the standard modal
decomposition of finite-dimensional linear systems
\cite[Sec.~2.5.2]{kailath1980linear}.  
In continuous time, modes associated with
system eigenvalues \(\lambda_r\) appear as exponentials \(e^{\lambda_r t}\). For the
discrete-time rational transfer function \(H(z)\), the analogous modal terms are powers
\(z_r^n\), where \(z_r\) are the poles of \(H(z)\). Equivalently, this follows from the
pole-residue expansion of the rational transfer function.

After separating the finite polynomial part of the transfer function, the
impulse response can be decomposed as
\begin{equation}
    h(n) = h_F(n) + h_R(n),
    \label{eq:pz-head-tail}
\end{equation}
where \(h_F(n)\) is a finite-length component and \(h_R(n)\) is the recursive modal tail.
For the improper case \(P_B > P_A\), the finite component has length \(P_F+1\), with
\(P_F = P_B-P_A\), and can be written as
\begin{equation}
    h_F(n)
    =
    \sum_{k=0}^{P_F} f_k \delta(n-k).
    \label{eq:pz-finite-head}
\end{equation}
For a proper transfer function, this polynomial part is absent; nevertheless, in the
acoustic interpretation below, we separately model the finite early-reflection component
\(h_F(n)\) as a sparse sum of delayed impulses.

Assuming simple poles \(z_r\), \(r=1,\ldots,P_A\), the recursive tail is a sum of modal
terms,
\begin{equation}
    h_R(n)
    =
    \begin{cases}
        0, & 0 \le n \le P_F,\\[1mm]
        \displaystyle
        \sum_{r=1}^{P_A}
        \beta^{(r)} z_r^{n-P_F-1},
        & n \ge P_F+1,
    \end{cases}
    \label{eq:pz-modal-tail}
\end{equation}
where \(\beta^{(r)}\in\complex\) are the pole residues. Since \(h(n)\in\reals\), non-real
poles and residues occur in complex-conjugate pairs. In the following, we use
\eqref{eq:pz-finite-head}--\eqref{eq:pz-modal-tail} as the starting point for showing that
the tensorized finite component and modal tail admit low-rank tensor representations.

\subsection{Tensor Representation}
\label{sec:joint_comp:tensor-representation}
Due to the infinite length of the tail, we further look at a truncated version of the
impulse response with length \(L\).  
Let us consider the head \(h_F(n)\) for \(n = 0, 1, \ldots, P_F\) which in practical terms
consists of the early reflections of the \gls{rir}.  
This part of the \gls{rir} for the typical small-to-medium sized room is often considered
on the order of 30 to 80 milliseconds, which corresponds to a few thousand samples
depending on the sampling rate (e.g., at 48 kHz, this is 1500 to 4000 samples).  
Because this part of the \gls{rir} consists of a handful \(N \ll P_F\) echoes,
\begin{equation*}
    h_F(n) = \sum_{k=1}^{N} \alpha_k \delta(n-\tau_k),
\end{equation*}
where \(\alpha_k\) are the amplitudes of the echoes and \(\tau_k \in \naturals\) are their
delays.  
Let there be the multi-index \(\tau_k = (i_1^{(k)}, i_2^{(k)}, \ldots,
i_D^{(k)})\) defined in \eqref{eq:multi-index}.  
We define the standard basis vectors
\begin{equation}
    \vect{e}^{(k, d)} = \left[0, \ldots, 0, 1, 0, \ldots, 0\right]^\T \in \reals^{I_d}
\end{equation}
with \(1\) at the \(i_d^{(k)}\)-th position.  
Then the impulses \(\delta(n-\tau_k)\) can be represented as rank-1 outer products 
with which we can write the tensorization of the head \(h_F(n)\) as
\begin{equation}
    \begin{aligned}
        \tens{H}_F &= \sum_{k=1}^{N} \alpha_k \bigl(
            \vect{e}^{(k, 1)} \circ \vect{e}^{(k, 2)} \circ \cdots \circ \vect{e}^{(k, D)}
        \bigr)\\
        &= \sum_{k=1}^{N} \alpha_k \nprod{d \in [D]} 
            \vect{e}^{(k, d)}
    \end{aligned}\label{eq:early-refl-tensorization}
\end{equation}
with \(\operatorname{rank}(\tens{H}_F) \le N\).\footnote{If early-reflection delays are
fractional, the delayed impulses can be represented by truncated interpolation kernels,
for example windowed sinc functions. Under the tensorization map, each non-zero sample of
such a kernel corresponds to one standard-basis rank-one term. Therefore, if the \(k\)-th
truncated kernel has \(T_k\) non-zero samples, the bound becomes
\(\operatorname{rank}(\tens{H}_F)\leq \min\{\sum_{k=1}^{N}T_k,P_F+1\}\).}
Note that here the cores of the \gls{btd} are scalar with \(K_d = 1\), i.e., a \gls{cp}.  


Then, since for the tail \(h_R(n)\) in \eqref{eq:pz-modal-tail},
\begin{equation}
    h_R(n) = \sum_{r=1}^{P_A} \beta^{(r)} z_r^{n-P_F-1}, \quad n \ge P_F+1,
\end{equation}
one finds that using the tensorization bijection \eqref{eq:multi-index} for any
\((i_{1},\dots,i_{D})\) we have
\begin{equation}
\begin{aligned}
    &\tens{H}_R [i_{1},\dots,i_{D}] =  \sum_{r=1}^{P_{A}}
    \beta^{(r)} z_{r}^{\left( i_{1} + i_{2} I_{1} + \cdots + i_{D} I_{1} I_{2} \cdots I_{D-1} - P_{F}-1\right)}\\
    &= \sum_{r=1}^{P_{A}}
    \widetilde{\beta}^{(r)}
    \left(z_{r}^{1}\right)^{i_{1}} 
    \left(z_{r}^{I_{1}}\right)^{i_{2}} 
    \left(z_{r}^{I_{1}I_{2}}\right)^{i_{3}} 
    \cdots 
    \left(z_{r}^{I_{1}I_{2}\cdots I_{D-1}}\right)^{i_{D}},
\end{aligned}
\end{equation}
where \(\widetilde{\beta}^{(r)} = \beta^{(r)}z_{r}^{-P_{F}-1}\).  
So equivalently, we can write the tensor \(\tens{H}_R\) as the sum of the outer products 
\begin{equation}
    \begin{aligned}
        \tens{H}_R &= \sum_{r=1}^{P_{A}}
        \widetilde{\beta}^{(r)} \left(
            \vect{a}^{(r, 1)} \circ \vect{a}^{(r, 2)} \circ \cdots \circ \vect{a}^{(r, D)}
        \right)\\
        &= \sum_{r=1}^{P_{A}} \widetilde{\beta}^{(r)} \nprod{d \in [D]} \vect{a}^{(r, d)}
    \end{aligned}\label{eq:pz-model-tail-tensor}
\end{equation}
which represents a complex-valued rank-\(P_{A}\) \gls{cp}, with the \(d\)th mode‐factor
\begin{equation}
    \label{eq:pz-model-factor}
    \vect{a}^{(r, d)}\left[i_{d}\right] = z_{r}^{s_{d} i_d}, \quad i_{d}=0,1,\dots,I_{d}-1
\end{equation}
for each \(r=1,\dots,P_{A}\) and \(d=1,\dots,D\), where \(s_d =
\prod_{\delta=1}^{d-1} I_{\delta}\) is the stride for mode \(d\).  
A complex-conjugate pole pair \(z_r=\rho_r e^{j \omega_r}\) and
\(z_r^{*} = \rho_r e^{-j\omega_r}\), with conjugate residues
\(\beta^{(r)}\) and \(\beta^{(r) \ast}\), can be written in real-valued form by expanding
the complex modal factors into damped cosine and sine components. Define
\begin{equation}
    \vect{c}_d[i_d]
    =
    \rho_r^{s_d i_d}\cos(\omega_r s_d i_d),
    \quad
    \vect{s}_d[i_d]
    =
    \rho_r^{s_d i_d}\sin(\omega_r s_d i_d).
\end{equation}
Since
\[
    \vect{a}^{(r,d)} = \vect{c}_d + j\vect{s}_d ,
\]
the outer product of the complex modal factors expands as
\begin{equation}
    \mybigcirc_{d=1}^{D}
    \left(\vect{c}_d+j\vect{s}_d\right)
    =
    \sum_{\mathcal S\subseteq[D]}
    j^{|\mathcal S|}
    \mybigcirc_{d=1}^{D}
    \vect{u}_d^{(\mathcal S)},
    \label{eq:complex-pair-expansion}
\end{equation}
where
\begin{equation}
    \vect{u}_d^{(\mathcal S)}
    =
    \begin{cases}
        \vect{s}_d, & d\in\mathcal S,\\
        \vect{c}_d, & d\notin\mathcal S.
    \end{cases}
\end{equation}
Adding the conjugate pole contribution gives the real-valued tensor
\begin{equation}
    \tens{H}_{R}^{(r)}
    =
    \sum_{\mathcal S\subseteq[D]}
    2\rho_r^{-P_F-1}
    \Ree{\beta^{(r)}j^{|\mathcal S|}}
    \mybigcirc_{d=1}^{D}
    \vect{u}_d^{(\mathcal S)} .
    \label{eq:real-cpd-conjugate-pair}
\end{equation}
Thus, each conjugate pole pair contributes at most \(2^D\) real-valued CPD rank-one
terms. Equivalently, by grouping \(\vect{c}_d\) and \(\vect{s}_d\) into
\(\matr{U}_d=[\vect{c}_d\ \vect{s}_d]\in\reals^{I_d\times 2}\), the same contribution can
be represented as one real-valued BTD block with multilinear ranks \(K_d=2\),
\(d=1,\ldots,D\), whose \(2\times\cdots\times2\) core contains the coefficients in
\eqref{eq:real-cpd-conjugate-pair}.
If some poles are real-valued, their contribution is a single real-valued rank-1 \gls{cp}
term each, which we omit in our representation without loss of generality.  

The tensorized \gls{rir} can therefore be written as the sum of the head and tail
\begin{equation}
    \label{eq:pz-model-tensorized}
    \tens{H} = \tens{H}_F + \tens{H}_R
\end{equation}
with \(\operatorname{rank}(\tens{H}) \le N + 2^{D-1} P_{A}\) in real-valued \gls{cp} form
and as a real-valued \gls{btd} consisting of \(N\) scalar-core blocks for the finite
component and \(P_A/2\) rank-\((2,\ldots,2)\) blocks for the conjugate pole pairs.  

\subsection{Factor Subspace Similarities}
\label{sec:joint_comp:shared-factor-subspaces}
Let \(m=1,\ldots,M\) index \glspl{rir} measured at different positions in the same room,
and assume that each tensorized \gls{rir} admits the representation
\(\tens{H}_m=\tens{H}_{mF}+\tens{H}_{mR}\) in
\eqref{eq:pz-model-tensorized}. For each mode \(d\), define the factor subspace
\begin{equation}
    \mathcal{S}_m^{(d)}
    =
    \operatorname{span}
    \left\{
        \vect{e}_m^{(1,d)},\ldots,\vect{e}_m^{(N,d)},
        \vect{a}_m^{(1,d)},\ldots,\vect{a}_m^{(P_A,d)}
    \right\}.
    \label{eq:subspace-span}
\end{equation}
Here, the vectors \(\vect{e}_m^{(k,d)}\) are associated with the finite early-reflection
component, while the vectors \(\vect{a}_m^{(r,d)}\) are associated with the modal tail.

Under the common-pole model for \glspl{rir} measured in the same room
\cite{hanedaCommonAcousticalPole1994}, the pole locations \(z_r\) are shared across
positions. Therefore, in the ideal modal-tail model, the factors
\(\vect{a}_m^{(r,d)}\) in \eqref{eq:pz-model-factor} are shared across \(m\), up to
scaling and ordering ambiguities. The early-reflection factors, in contrast, are
generally position-dependent. Consequently, the subspaces
\(\mathcal{S}_m^{(d)}\) are not expected to be identical across positions, but may exhibit
similarities whenever common modal components, similar reflection delays, or other
position-dependent acoustic regularities lead to aligned factor subspaces.

\begin{figure}[t]
    \centering
    \includegraphics[width=\columnwidth]{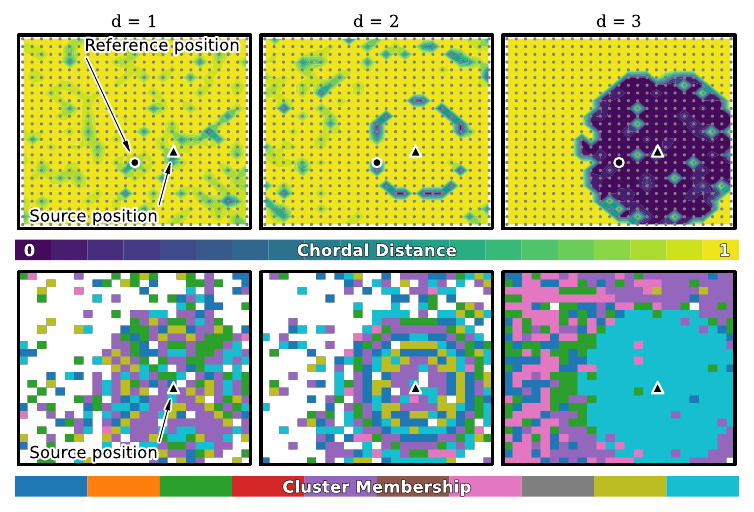}
    \caption[Chordal distance and spectral clustering illustration]{
        \textbf{Top row:} Top views of a shoebox room with the Chordal distance between 
        tensorized \glspl{rir} at a reference position and a grid of positions over the
        room.  
        The distances are computed from the factor subspaces in mode \(d=1,2,3\), respectively.  
        Darker colors indicate higher similarity (lower distance).  
        \textbf{Bottom row:} Top views of the same room with spectral clustering applied.  
        The clusters are indicated by different colors and show the spatial arrangement of the
        microphone positions within each cluster.  
    }
    \label{fig:joint_comp:grassmann-distance}
\end{figure}

We quantify these similarities by comparing subspaces across positions. Let
\(d_S(\mathcal{S}_i^{(d)},\mathcal{S}_j^{(d)})\) denote a standard subspace distance, such
as the chordal or Grassmannian distance \cite{yeSchubertVarietiesDistances2016}.
\autoref{fig:joint_comp:grassmann-distance} shows the resulting distances for \glspl{rir}
simulated on a grid in a shoebox room.  
The observed patterns indicate that the factor
subspaces contain non-trivial spatial structure.  
While part of this structure is
consistent with the common-pole interpretation of the late reverberant tail, some patterns
may also reflect early-reflection or delay-related structure induced by the room geometry
and the tensorization.  
A detailed separation of these effects is beyond the scope of this
work, but the observation motivates the data-driven sharing strategy used in the proposed
\gls{jmcbtd}: rather than imposing a fixed sharing pattern a priori, we infer shared
factors from factor-subspace affinity graphs.

\section{Joint Multichannel RIR Compression using BTD}
\label{sec:joint_comp:joint-multichannel-room-impulse-response-compression-using-btd}
Based on the theoretical background of the shared factors subspaces, we can now propose a
joint compression framework for multichannel \glspl{rir} using the \gls{btd}.  
We first introduce the \gls{sbtd} and then the \gls{jmcbtd}.  

\subsection{Stacked Multichannel BTD (SMCBTD)}
\label{sec:joint_comp:joint-multichannel-room-impulse-response-compression-using-sbtd}
We stack the \(M\) tensorized \glspl{rir} \(\tens{H}_m\) as defined in
\eqref{eq:rir-tensorization} into a \((D\!+\!1)\)-mode tensor \(\tens{T} \in \reals^{
I_1 \times I_2 \times \cdots \times I_D \times M}\) such that
\begin{equation}
    \tens{T}[i_1, i_2, \ldots, i_D, m] = \tens{H}_m[i_1, i_2, \ldots, i_D].
\end{equation}
Then we can apply \gls{btd} \eqref{eq:btd-minimization} to the tensor \(\tens{T}\) such
that 
\begin{equation}
    \tens{T} \approx \widehat{\tens{T}} =
        \sum_{r=1}^{R} \tens{G}^{(r)}
        \nprod{d \in [D+1]} \matr{A}^{(r, d)}
\end{equation}
with the analogous definitions as in \autoref{sec:joint_comp:btd} for a \((D\!+\!1)\)-mode tensor.  
From \autoref{sec:joint_comp:shared-factor-subspaces}, we can expect
that the added channel mode captures inter-\gls{rir} redundancy, while the remaining modes
capture the intra-\gls{rir} low rank.  
The compression ratio for the \gls{sbtd} is defined as
\begin{equation}
    \label{eq:compression-ratio-sbtd}
    C_{\text{\gls{sbtd}}} = 1 - \frac{
        \sum_{r=1}^{R} N_r 
    }{M L}
\end{equation}
with 
\begin{equation}
    N_r = \prod_{d=1}^{D+1} K_d^{(r)} + \sum_{d=1}^{D} I_d K_d^{(r)} + M K_{D+1}^{(r)}.
\end{equation}
With high channel count \(M\), large \gls{rir} length \(L\), and large number of blocks
\(R\), the \gls{sbtd} becomes computationally expensive and (temporarily) memory
intensive.  
To alleviate this, we can consider a per-\gls{rir} decomposition with shared factors
across the \glspl{rir} as described in the next section.

\subsection{Joint Multichannel BTD (JMCBTD)}
\label{sec:joint_comp:joint-multichannel-room-impulse-response-compression-using-jmcbtd}
In addition to the \gls{sbtd}, we can formulate a joint compression framework for
multichannel \glspl{rir} using the \gls{jmcbtd} with spectral clustering for improved factor
sharing and compression efficiency.  
The similarity of the factor subspaces across \glspl{rir} implies that a reasonable
approximation could be made by shared factors across \glspl{rir}, and thus reducing the
number of parameters.  
For this, for each \((r,d) \in \set{I}\), we need to define clusters of \glspl{rir}
that share a factor matrix.  
Let such a cluster be denoted by
\begin{equation}
    \set{C}_c^{(r,d)} \subseteq \{1, \ldots, M\}
\end{equation}
where \(c = 1,\ldots,Q^{(r,d)}\) is the cluster index and
\(Q^{(r,d)}\) the number of clusters for the specific \((r,d)\).  
The clusters are disjoint, i.e., \(\set{C}_c^{(r,d)} \cap \set{C}_{c'}^{(r,d)} = \emptyset\) for
\(c \neq c'\) and all \((r, d)\).  
Define the factor matrices as
\begin{equation}
    \matr{A}_m^{(r, d)} = \begin{cases}
        \matr{U}_c^{(r, d)} & \text{if } m \in \set{C}_c^{(r, d)},\\
        \matr{U}_m^{(r, d)} & \text{otherwise,}
    \end{cases} 
\end{equation}
where \(\matr{U}_c^{(r, d)}\) denote the factors shared for cluster \(\set{C}_c^{(r, d)}\)
and \(\matr{U}_m^{(r, d)}\) the individual factors for each \(m\).  
With this, we define the optimization problem for the joint multichannel \gls{rir}
compression as
\begin{mini}
    {}{\sum_{m=1}^{M} \Bigl\|
        \tens{H}_m - \sum_{r=1}^{R} \tens{G}_m^{(r)} \nprod{d \in [D]} \matr{A}_m^{(r, d)}
    \Bigr\|_F^2}
    {\label{eq:joint-multichannel-btd-minimization}}{}
    \addConstraint{
        \matr{A}_m^{(r, d) \T} \matr{A}_m^{(r, d)}
    }{= \matr{I}}{\quad m\!= \!1, \ldots, M, \, (r,d) \in \set{I}}
\end{mini}
As in the single-channel problem, the orthogonality constraint allows the application of
an \gls{hooi} approach, again, limiting the multilinear core dimensions (multilinear rank)
to \(K_d^{(r)} \leq I_d\) for all \(d\) and \(r\).  
The shared and individual factors are treated separately.  
While the update step for individual factors is unchanged, the shared factor update is
performed using the \gls{svd} on stacked aggregated residuals from within the cluster
such that 
\begin{equation}
    \matr{U}_c^{(r, d)} = \lsv{K_d^{(r)}}{
        \bar{\matr{Y}}_{c (d)}^{(r)}
    },
\end{equation}
where 
\begin{equation}
    \bar{\matr{Y}}_{c (d)}^{(r)} =
        \big[
            \matr{Y}_{m (d)}^{(r)}
        \big]_{m \in \set{C}_c^{(r, d)}}
    \label{eq:per-cluster-stack}
\end{equation}
is the horizontal concatenation of the residual unfoldings for all \glspl{rir} \(m\) in the
cluster \(\set{C}_c^{(r, d)}\).  
The core tensors are updated as before.  
\autoref{alg:joint-btd-hooi} summarizes the overall procedure.  
Since the choice of the clusters is non-trivial, we now introduce a two-step approach
using spectral clustering to determine data-driven sharing clusters in the next section.  

\begin{algorithm}
\caption[Alternating HOOI for JMCBTD]{Alternating HOOI for JMCBTD}
\label{alg:joint-btd-hooi}
\begin{algorithmic}[1]
\State \textbf{Input:}
    Tensors \(\tens{H}_m \in \reals^{I_1 \times I_2 \times \ldots \times I_D}\),
    number of blocks \(R\),
    block ranks \((K_1^{(r)}, K_2^{(r)}, \ldots, K_D^{(r)})\),
    clusters \(\set{C}_c^{(r,d)}\),
    outer iteration count \(N_{\text{o}}\),
    inner iteration count \(N_{\text{i}}\)
\State \textbf{Output:}
    Core tensors \(\tens{G}_m^{(r)} \in \reals^{K_1^{(r)} \times K_2^{(r)} \times \ldots
    \times K_D^{(r)}}\)
    , factor matrices
    \(\matr{U}_m^{(r, d)} \in \reals^{I_d \times K_d^{(r)}}\), 
    \({\matr{U}}_c^{(r, d)} \in \reals^{I_d \times K_d^{(r)}}\) 
\State \textbf{Initialize:} Random \(\matr{U}_m^{(r, d)}\) for each \(r, d, m\),
random \({\matr{U}}_c^{(r, d)}\) for each \(r, d, c\),
random \(\tens{G}_m^{(r)}\) for each \(r, m\)
\For{\(N_{\text{o}}\)}
    \For{\(r = 1, \ldots, R\)}
        \For{\(m=1, \ldots, M\)}
            \State \(\tens{R}_m^{(r)} = \tens{H}_m - \sum_{l \neq r}
                \widehat{\tens{H}}_m^{(l)}\)
        \EndFor
        \For{\(N_{\text{i}}\)}
            \For{\(d = 1, \ldots, D\)}
                \For{\(m=1, \ldots, M\)}
                    \State 
                    \(
                        \matr{Y}_{m (d)}^{(r)} \gets \tens{R}_m^{(r)} 
                            \nprod{\delta \neq d} \matr{A}_m^{(r, \delta) \T}
                    \)
                    \If{\(m \notin \bigcup_{c=1}^{Q^{(r,d)}} \set{C}_c^{(r, d)}\)}
                        \State \(
                            \matr{U}_m^{(r, d)} \! \gets \lsv{K_d^{(r)}}{\matr{Y}_{m (d)}^{(r)}}
                        \)
                    \EndIf
                \EndFor
                \For{\(c = 1, \ldots, Q^{(r,d)}\)}
                    \State \(\bar{\matr{Y}}_{c (d)}^{(r)} =
                            \big[
                                \matr{Y}_{m (d)}^{(r)}
                            \big]_{m \in \set{C}_c^{(r, d)}}\)
                    \State
                    \(
                        {\matr{U}}_c^{(r, d)} \!\gets \lsv{K_d^{(r)}}{
                            \bar{\matr{Y}}_{c (d)}^{(r)}
                        }
                    \)
                \EndFor
            \EndFor
        \EndFor
        \For{\(m=1, \ldots, M\)}
            \State
            \(
                \tens{G}_m^{(r)} \gets
                \tens{H}_m \nprod{d \in [D]} \matr{A}_m^{(r, d) \T}
            \)
        \EndFor
    \EndFor
\EndFor
\end{algorithmic}
\end{algorithm}

\subsection{Spectral Clustering on Factor Subspace Affinity Graphs}
\label{sec:joint_comp:spectral-clustering}
The choice of channel clusters \(\set{C}_c^{(r,d)}\) is crucial for the performance of the
proposed method.  
A fixed sharing pattern could be imposed manually, for example by sharing selected
modes across all channels.  
Instead, we use a data-driven heuristic based on spectral clustering of factor-subspace
affinity graphs.  
We define the factor subspace affinity graph for a \((r,d)\) as 
\begin{equation}
    \set{G}^{(r, d)} \triangleq \{\set{V}^{(r, d)}, \set{E}^{(r, d)}, w^{(r, d)}\}
\end{equation}
where
\(\set{V}^{(r, d)} = \{\matr{U}_1^{(r, d)}, \ldots, \matr{U}_M^{(r, d)}\}\)
is the set of factors for all \glspl{rir}, \(\set{E}^{(r, d)} = \{(i,j) \mid
w^{(r,d)}(i,j) > \tau\}\) is the set of edges, and \(w^{(r,d)}\) is the weight function.  
The threshold \(\tau \geq 0\) prunes the fully connected graph to a significant
subgraph.   
The weight function \(w^{(r,d)}(i,j)\) is defined as an affinity value based on the
subspace distance by 
\begin{equation}
    w^{(r,d)}(i,j) = \exp \Bigl(
        -\frac{d_{S}(\matr{U}_i^{(r, d)}, \matr{U}_j^{(r, d)})}{\sigma^2}
    \Bigr), 
\end{equation}
where \(d_{S}(\cdot,\cdot)\) is the Chordal distance and \(\sigma\) is a scaling
factor.  
Note that we treat the factor matrices \(\matr{U}_m^{(r, d)}\) as bases of the subspaces
due to the orthogonality constraint.  

To find the clusters, we employ spectral clustering on the graphs
\(\set{G}^{(r,d)}\).  
The affinity matrix \(\matr{W}^{(r,d)} \in \reals^{M \times M}\) is constructed from the
weight function \(w^{(r,d)}(i,j)\) as follows:
\begin{equation}
    \matr{W}^{(r,d)}[i,j] = \begin{cases}
        w^{(r,d)}(i,j) & \text{if } (i,j) \in \set{E}^{(r,d)} \\
        0 & \text{otherwise}
    \end{cases}
\end{equation}
We then truncate all affinity values below the threshold \(\tau\) and remove the isolated
nodes from the graph, i.e., rows and columns of \(\matr{W}^{(r,d)}\) where the row and
column-wise sum of elements is 0.  
From the truncated affinity matrix, denoted \(\tilde{\matr{W}}^{(r,d)} \in
\reals^{\tilde{M} \times \tilde{M}}\) with \(\tilde{M} \leq M\), we compute the normalized
Laplacian matrix
\begin{equation}
    \matr{L}^{(r,d)} = \matr{I} -
        \bigl(\matr{D}^{(r,d)}\bigr)^{-1/2}
            \tilde{\matr{W}}^{(r,d)}
        \bigl(\matr{D}^{(r,d)}\bigr)^{-1/2}
\end{equation}
where the diagonal matrix \(\matr{D}^{(r,d)}\) is the degree matrix defined by
\(\matr{D}^{(r,d)}[i,i] = \sum_{j} \tilde{\matr{W}}^{(r,d)}[i,j]\).  
We then compute the eigenpairs \((\lambda_n^{(r,d)}, \vect{v}_n^{(r,d)})\) of the
Laplacian matrix \(\matr{L}^{(r,d)}\) using \gls{evd}, then select the \(Q^{(r,d)}\) most
significant eigenvalues and their corresponding eigenvectors through the maximum eigengap
of the sorted sequence \(\lambda_1^{(r,d)} \leq \lambda_2^{(r,d)} \leq \cdots \leq
\lambda_{\tilde{M}}^{(r,d)}\) such that
\begin{equation}
    Q^{(r,d)} = \min\{ \arg\!\max_{1 \leq i < \tilde{M}} \bigl(\lambda_{i+1}^{(r,d)} - \lambda_{i}^{(r,d)}\bigr), Q_{\text{max}}\},
\end{equation}
where \(Q_{\text{max}}\) is a safeguard maximum number of clusters.  
After row-normalization of the matrix formed by the eigenvectors
\begin{equation}
    \tilde{\matr{V}}^{(r,d)} = \begin{bmatrix}
        \vect{v}_1^{(r,d)} &
        \cdots &
        \vect{v}_{Q^{(r,d)}}^{(r,d)}
    \end{bmatrix},
\end{equation}
we apply \(k\)-means clustering on the rows to obtain the final clusters
\(\set{C}_c^{(r,d)}\) for \(c = 1,\ldots,Q^{(r,d)}\).  
This partitions the \(M\) \glspl{rir} into \(Q^{(r,d)}\) clusters, i.e., subgraphs of
\(\set{G}^{(r,d)}\), for each \((r,d)\) based on their similarity in the factor subspaces,
and \(M - \tilde{M}\) individual \glspl{rir} that do not belong to any cluster.    
While there is still some freedom in choosing the parameters \(\tau\), \(\sigma\), and
\(C_{\text{max}}\), this approach is a more systematic data-driven method than manual
choice of clusters.  
These clusters can then be used to initiate the joint \gls{jmcbtd} compression algorithm.  
The overall procedure is summarized in \autoref{alg:two-step-compression}.  
\begin{algorithm}[t]
\caption[Two-step compression procedure]{\gls{jmcbtd} using spectral clustering}
\label{alg:two-step-compression}
\begin{algorithmic}[1]
\For{\(m = 1,\ldots,M\)}
    \State Compute individual \glspl{btd} for \(\tens{H}_m\)
\EndFor
\For{\((r, d) \in \set{I}\)}
    \State Construct graph
    \(\set{G}^{(r,d)}\)
    \State Apply spectral clustering to obtain
    \(\set{C}_c^{(r,d)}\)
\EndFor
\State Compute joint \gls{btd} (\autoref{alg:joint-btd-hooi}) using clusters \(\set{C}_c^{(r,d)}\)
\end{algorithmic}
\end{algorithm}
The compression ratio for the \gls{jmcbtd} is defined as
\begin{equation}
    \label{eq:compression-ratio-jmcbtd}
    C_{\text{\gls{jmcbtd}}} = 1 - \frac{
        N_{\text{cores}} +
        \sum_{(r,d) \in \set{I}} \Bigl(
            N^{(r,d)}_\text{shr} + N^{(r,d)}_\text{ind}
        \Bigr)
    }{M L}
\end{equation}
where 
\begin{align}
    N_{\text{cores}} &= M \sum_{r=1}^{R} \prod_{d=1}^{D} K_d^{(r)},\\
    N^{(r,d)}_{\text{shr}} &= Q^{(r,d)} I_d K_d^{(r)}, \text{and}\\
    N^{(r,d)}_{\text{ind}} &= \sum_{m \notin \bigcup_{c=1}^{Q^{(r,d)}} \set{C}_c^{(r,d)}} I_d K_d^{(r)}
\end{align}
are the number of coefficients for the cores, shared factors, and individual factors,
respectively.  

Note that since the order of blocks \(r\) is arbitrary in the \gls{btd}, and therefore
block contributions \((r, d)\) across different channels \(m\) might not represent the
same characteristics, it is advantageous to reorder them such that the subspace distances
are minimized before clustering.  
This can be done for example using the Hungarian (Kuhn–Munkres) algorithm
\cite{kuhn1955hungarian,munkres1957algorithms} or the Auction algorithm
\cite{bertsekas1979distributed}.  
We apply the latter using the Chordal distance as the cost metric, such that
similar blocks are grouped together for better clustering and compression performance.  

\subsection{Computational Complexity and Memory Usage}
\label{sec:joint_comp:computational-complexity}
Define \(J^{(r)}_d=\prod_{\delta\neq d} K^{(r)}_\delta\).  
We use \(W\) for "work" to denote cost here, not to confuse with the compression ratio.  
One outer iteration with \(N_i\) inner iterations has cost 
\begin{equation}
    \sum_{r=1}^{R}\!\Big[
        W^{(r)}_{\text{rec}} + N_{\text{i}}\!\sum_{d=1}^{D}\!\big(
            W^{(r,d)}_{Y}+W^{(r,d)}_{\mathrm{SVD}}
        \big) + W^{(r)}_{\mathrm{core}} \Big],
\end{equation}
where we use standard multi-mode product \cite{koldaTensorDecompositionsApplications2009}
and truncated-\gls{svd} costs \cite{golubMatrixComputations2013},
\begin{align}
    W^{(r)}_{\text{rec}} &= \sum_{d=1}^{D} I_d K^{(r)}_d \big(\prod_{\delta < d} I_\delta \big) \big( \prod_{\delta>d} K^{(r)}_\delta \big) ,\\
    W^{(r,d)}_{Y} &=
        \sum_{\delta = 1}^{d-1} I_\delta^{} K^{(r)}_\delta \big( \prod_{\mu = 1}^{\delta-1} K^{(r)}_\mu \big) 
        \big(  \!  \! \prod_{\mu = \delta + 1}^{d-1}  \!  \! I_\mu \big) I_d \big(  \! \! \prod_{\mu = d+1}^{D}  \!  \! I_\mu \big)\nonumber \\
        + &\sum_{\delta = d+1}^{D} I_\delta^{} K^{(r)}_\delta \big( \prod_{\mu = 1}^{d-1}  K^{(r)}_\mu \big) 
        I_d \big(  \!  \! \prod_{\mu = d+1}^{\delta-1}  \!  \! I_\mu \big) \big(  \!  \! \prod_{\mu = \delta + 1}^{D}  \!  \! K^{(r)}_\mu \big)
        ,\\
    W^{(r,d)}_{\mathrm{SVD}} &= I_d J^{(r)}_d K^{(r)}_d,\\
    W^{(r)}_{\mathrm{core}} &= \sum_{d=1}^{D} I_d K^{(r)}_d \big(\prod_{\delta < d} K^{(r)}_\delta \big) \big( \prod_{\delta>d} I_\delta \big).
\end{align}
While these expressions for the exact costs seem complex, their intuition is rather
straightforward: for the reconstruction cost \(W^{(r)}_{\text{rec}}\), the sequential
multi-mode products over the sequence of modes \(d\), i.e., a matrix vector product for
each fiber, where the number of fibers change after each mode from \(I_\delta\) to
\(K^{(r)}_\delta\), and the same applies for the \(W^{(r)}_{\mathrm{core}}\) cost in
reverse.  
The cost \(W^{(r,d)}_Y\) is a similar multi-mode product but skipping mode \(d\), which
separates the expression into two sums.  
Peak working memory is 
\(
    L +\max_{r,d} I_d\,J^{(r)}_d.
\)

Stacking \(M\) \glspl{rir} for \gls{sbtd} forms a \((D \! + \! 1)\)-mode tensor with a
channel mode of size \(M\) and applies \gls{btd} (\autoref{sec:joint_comp:btd}).  
The channel-mode update performs a truncated \gls{svd} of size \(M \times \prod_{d=1}^{D}
K_d^{(r)}\), and all contractions traverse the full \(ML\) entries per each sweep.  
\gls{jmcbtd} instead updates shared factors on per-cluster stacks
\eqref{eq:per-cluster-stack}, where the dominant \gls{svd} in mode \(d\) has size \(I_d
\times \sum_{m \in \set{C}_c^{(r,d)}} J_d^{(r)}\), replacing
\(M\) by the cluster size \(|\set{C}_c^{(r,d)}|\).  
Thus, the sweeps scale with \(\max_c | \set{C}_c^{(r,d)} |\).  
The peak memory shrinks accordingly.  
When \(\max_c | \set{C}_c^{(r,d)} | \ll M\) and \(L\) is large, \gls{jmcbtd} reduces both
computational cost and memory usage and the one-off cost of \(M\) individual \glspl{btd}
and spectral clustering becomes negligible.  
This was confirmed by numerical experiments in \autoref{sec:joint_comp:numerical-simulations}, where
\gls{jmcbtd} achieved faster computation times than \gls{sbtd}, and at larger \(M\) and
\(L\) the latter became infeasible due to memory constraints.  
By caching residuals, one can trade memory for computational effort (time), which becomes
relevant for large datasets and when memory is the limiting factor.

\section{Numerical Simulations}
\label{sec:joint_comp:numerical-simulations}
We conduct numerical experiments to evaluate the performance of the proposed method.  
For this, we first use simulated \glspl{rir} in shoebox rooms to show spatial clustering
effects and subsequently use measured \glspl{rir} from the \gls{smard}
\cite{nielsenSingleMultichannelAudio2014}
and \gls{sriracha} \cite{pellingSRIRACHAShoeboxRoom2025}
and \gls{churchir} dataset \cite{giampiccolo2025churchir}
to validate the applicability to real-world scenarios.  
Let the measure for the reconstruction error be the \gls{nm} defined as
\begin{equation}
        \text{NM}
            = \frac{\| \tens{H}_m - \widehat{\tens{H}}_m \|_F}{\| \tens{H}_m \|_F}
            = \frac{\| \h_m - \hat{\h}_m \|_2}{\| \h_m \|_2},
            \label{eq:joint_comp:normalized-misalignment}
\end{equation}
where \(\tens{H}_m\) is the original tensor, \(\widehat{\tens{H}}_m\) is the reconstructed
tensor, \(\h_m\) is the original vectorized \gls{rir} and \(\hat{\h}_m\) is the
reconstructed vectorized \gls{rir}.  
Let the mean \gls{nm} be defined as
\begin{equation}
        \overline{\text{NM}}
            = \frac{1}{M} \sum_{m=1}^{M} \text{NM},
\end{equation}
and let
\begin{equation}
    \overline{\text{NM}}_{\text{dB}} = 20 \log_{10}(\overline{\text{NM}}) \quad [\mathrm{dB}]
\end{equation}
be the \gls{nm} in decibel scale.  

For any set of \glspl{rir}, be it simulated or measured, we first estimate the
reverberation time \(T_{60}\) and then truncate the \glspl{rir} to \(L \geq L_{T_{60}}
= T_{60} f_s\), where we define this as the smallest number of samples such that we have
the most equidimensional tensor shape.  
For this, let \(u = \lfloor \sqrt[D]{L_{T_{60}}}\rfloor\) be the largest integer such
that \(u^D \leq L_{T_{60}}\).  
Then choose the smallest \(v \in \{1,...,D\}\) such that \(u^{D-v} (u+1)^v \geq
L_{T_{60}}\).  
If necessary, we zero-pad the \glspl{rir} to length \(L\).  
So then, we set the tensor dimensions as
\begin{equation}
    I_d = \begin{cases}
        u+1 & \text{if } d \leq v,\\
        u & \text{otherwise}
    \end{cases}
\end{equation}
for \(d = 1,...,D\).  
We aim for an equidimensional tensor shape, due to the better compression performance
as discussed in \cite{jalmbyCompressionRoomImpulse2024}.  

The methods in our comparison are the proposed \gls{sbtd} and \gls{jmcbtd} as well as the
\gls{glram} method \cite{jalmbyMultiChannelLowRankConvolution2024} as a baseline.  
The \gls{glram} method is essentially a 2-mode Tucker decomposition, i.e., \gls{btd} with
\(R=1\).  
In this comparison, we consider the left matrices of the \gls{glram} as common and do a
refinement step with individual right matrices.  
The compression ratio for the \gls{glram} method is varied by changing the dimensions of
the core matrix.  
\begin{table*}
    \small
    \setlength{\tabcolsep}{4pt} 
    \renewcommand{\arraystretch}{1.25}
    \centering

    \begin{tabularx}{\textwidth}{|l|*{5}{>{\centering\arraybackslash}X|}}
        \hline
        \multirow{2}{*}{\textbf{Parameter}}
            & \multicolumn{2}{c|}{\textbf{Simulated Array}}
            & \textbf{SMARD 2X0X}
            & \textbf{SRIRACHA C1}
            & \textbf{ChurchIR} \\
        \cline{2-3}
            & \textbf{Far} & \textbf{Near} & & & \\
        \hline

        \multicolumn{6}{|c|}{\textbf{Room parameters}} \\
        \hline
        Size $[\ell,w,h]$
            & 8.00~m 8.00~m 3.00~m
            & 8.00~m 8.00~m 3.00~m
            & 2.87~m 7.34~m 8.09~m
            & 6.22~m 3.85~m 3.07~m
            & 51.08~m 24.89~m 23~m \\
        \hline
        \(T_{60}\)
            & \(0.53\)~s
            & \(0.53\)~s
            & \(0.15\)~s
            & \(0.5\)~s
            & \(T_{30} = 7~\mathrm{s~(trunc.)}\) \\
        \hline
        Critical distance \(d_c\)
            & \(1.12\)~m
            & \(1.12\)~m
            & \(1.92\)~m
            & \(0.69\)~m
            & \(2.81\)~m \\
        \hline
        Mixing time \(t_{\text{mix}}\)
            & \(14\)~ms
            & \(14\)~ms
            & \(13\)~ms
            & \(9\)~ms
            & \(171\)~ms \\
        \hline
        Source pos. $[\ell,w,h]$
            & 4.0~m 4.0~m 1.2~m
            & 1.0~m 1.0~m 1.2~m
            & see \cite{nielsenSingleMultichannelAudio2014}
            & see \cite{pellingSRIRACHAShoeboxRoom2025}
            & see \cite{giampiccolo2025churchir} \\
        \hline
        \(M\)
            & \(64\) & \(64\) & \(168\) & \(256\) & \(64\) \\
        \hline
        \(f_s\)
            & \(48\)~kHz & \(48\)~kHz & \(48\)~kHz & \(36\)~kHz & \(36\)~kHz \\
        \hline

        \multicolumn{6}{|c|}{\textbf{Decomposition parameters}} \\
        \hline
        \(D\)
            & \(3,4,5\) & \(3,4,5\) & \(3,4,5\) & \(3,4,5\) & \(3,4,5\) \\
        \hline
        \(K_d\)
            & \(1,2,4\) & \(1,2,4\) & \(1,2,4\) & \(1,2,4\) & \(1,2\) \\
        \hline
        \(\sigma\)
            & \(0.2\ldots 1.2\)
            & \(0.2\ldots 1.2\)
            & \(0.6\ldots 1.8\)
            & \(0.4\ldots 6.0\)
            & \(0.4\ldots 0.6\) \\
        \hline
        \(\tau\)
            & \(0.4\) & \(0.4\) & \(0.4\) & \(0.4\) & \(0.4\) \\
        \hline
        \(Q_{\text{max}}\)
            & \(6\) & \(6\) & \(6\) & \(6\) & \(6\) \\
        \hline
        \(N_{\text{o}}\)
            & \(80\) & \(80\) & \(80\) & \(80\) & \(80\) \\
        \hline
        \(N_{\text{i}}\)
            & \(8\) & \(8\) & \(8\) & \(8\) & \(8\) \\
        \hline
    \end{tabularx}
    \caption[Room and decomposition parameters for numerical experiments]{ Room and
        decomposition parameters for the numerical experiments. For the measured datasets,
        some parameters are referenced to the original publications. The critical distance
        is approximated by \(d_c = 0.057 \sqrt{V/T_{60}} = 1.12\)~m and the mixing time in
        ms is approximated by \(t_{\text{mix}} \approx \sqrt{V}\)
        \cite{defrance2008measuring}, where \(V\) is the room volume.  
    }
    \label{tab:parameters-experiments}
\end{table*}

\subsection{Simulated Room Impulse Responses}
\label{sec:joint_comp:simulated-room-impulse-responses}
\begin{figure}
    \centering
    \hspace{-0.4cm}
    \includegraphics[trim={0 0 0 0},clip,width=\columnwidth]{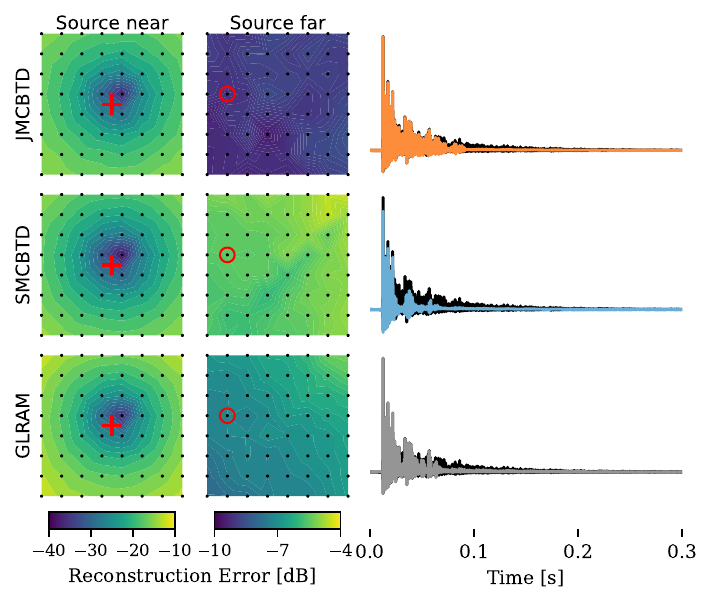}
    \caption[Reconstruction error over the array and examples of reconstructed RIRs.]{  
        This figure illustrates the reconstruction error \gls{nm} over the compact
        microphone array geometry (dots indicate microphones) for the two source
        positions.  
        For the near source position (left), it is marked \textcolor{red}{+}.  
        We observe the reconstruction error for the proposed \gls{jmcbtd} method
        is significantly lower across the array at the far-field position.  
        Additionally, example \glspl{rir} (black) and their reconstructions
        comparing \gls{jmcbtd} and \gls{sbtd} with tensorization order \(D=4\), and
        \gls{glram}, all at compression rate \(C=0.8\) for a selected microphone position
        (circled in red \(\textcolor{red}{\bm{\circ}}\)) are shown for the far-field
        position.  
    }
    \label{fig:joint_comp:spatial_error}
\end{figure}

We simulate a shoebox room using the \gls{rism}
\cite{desenaModelingRectangularGeometries2015}, placing a compact \(M = 8 \times 8 = 64\)
square grid microphone array with a grid spacing of \(0.1\)~m centered in the room of
dimensions \(8.00\text{~m} \times 8.00\text{~m} \times 3.00\text{~m}\).  
The absorption coefficients of the walls are set to achieve an average reverberation time
of \(T_{60} \approx 0.5\)~s.  
Two source positions are considered: a close position within the array's near field and a
far position in the array's far field, as illustrated in \autoref{fig:joint_comp:spatial_error}.  
The room dimensions and reverberation time are listed in \autoref{tab:parameters-experiments}.  
The \glspl{rir} are simulated at a sampling frequency of \(f_s = 48\)~kHz and truncated as
described above.  
We compare the proposed methods \gls{jmcbtd} and \gls{sbtd} with the \gls{glram}-based
method in \cite{jalmbyMultiChannelLowRankConvolution2024}.  
We vary the tensorization order \(D\) and the compression rate to evaluate the
performance.  
The decomposition parameters used are listed in \autoref{tab:parameters-experiments}.  

\begin{figure}
    \centering
    \hspace{-0.4cm}
    \includegraphics[width=\columnwidth]{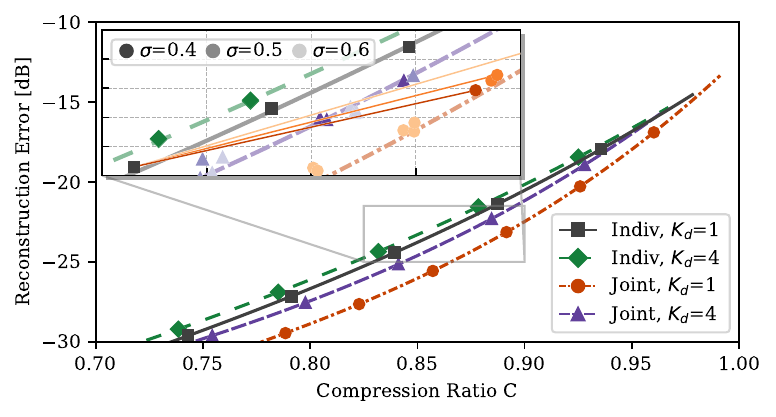}
    \caption[Joint vs.\ individual compression performance.]{ This figure shows the mean reconstruction error \gls{nm} versus compression
        rate comparing individual \gls{btd} and the proposed \gls{jmcbtd} for
        different core ranks \(K_d \in \{1, 4\}\) and clustering parameters \(\sigma \in
        \{0.4,0.5,0.6\}\).  
        The data is from the simulated compact microphone array at the near-field
        position, and tensorization order \(D=3\).  
        It shows that by clustering decomposition factors and sharing them across multiple
        channels, yields lower reconstruction error at the same compression rate.  
        We can see that the performance gain is more pronounced for lower core ranks,
        especially \(K_d = 1\).  
        Curves are cubic spline fits on the data for readability. In the inset figure, we
        indicate the effect of varying \(\sigma\) by drawing connecting lines between data
        points with the same block number \(R\) compressed individually and jointly. }
    \label{fig:joint_comp:joint-vs-individual}
\end{figure}

\autoref{fig:joint_comp:spatial_error} shows how the reconstruction error varies spatially across the
array for a specific compression rate \(C=0.8\) and tensorization order \(D=4\).  
We can observe that in this scenario, while all compared methods perform similarly for the
near-field position at a minimum reconstruction error at the closest microphone, the
advantage of the proposed \gls{jmcbtd} method is significant in the far-field scenario.  
We can observe the error increasing with distance from the center of the array for all
methods, but the \gls{jmcbtd} method maintains a lower error rate compared to the others
with \(-16\)~dB mean \gls{nm} compared to \(-13\)~dB for \gls{glram} and \(-12\)~dB for
\gls{sbtd} at the far-field position.  
The \gls{sbtd} shows similar performance to the \gls{glram}-based method in
\cite{jalmbyMultiChannelLowRankConvolution2024} in this scenario.  
Further, \autoref{fig:joint_comp:spatial_error} shows example \glspl{rir} and their reconstructions
for the far-field position for a compression rate of \(C=0.8\) comparing the proposed
\gls{jmcbtd} and \gls{sbtd} methods with the \gls{glram}-based method with tensorization
order \(D=4\).  

\autoref{fig:joint_comp:joint-vs-individual} shows the mean reconstruction error versus compression
ratio for the simulated compact microphone array at the near-field position with tensorization
order \(D=3\).  
We compare the proposed joint \gls{jmcbtd} with individual \gls{btd} for core ranks \(K_d
\in \{1, 4\}\) and clustering parameters \(\sigma \in \{0.4,0.5,0.6\}\).  
We can observe that by clustering decomposition factors and sharing them across multiple
channels, yields lower reconstruction error at the same compression rate.  
We can see that the performance gain is more pronounced for lower core ranks, especially
\(K_d = 1\).  

\begin{figure*}
    \centering
    \includegraphics[width=\textwidth]{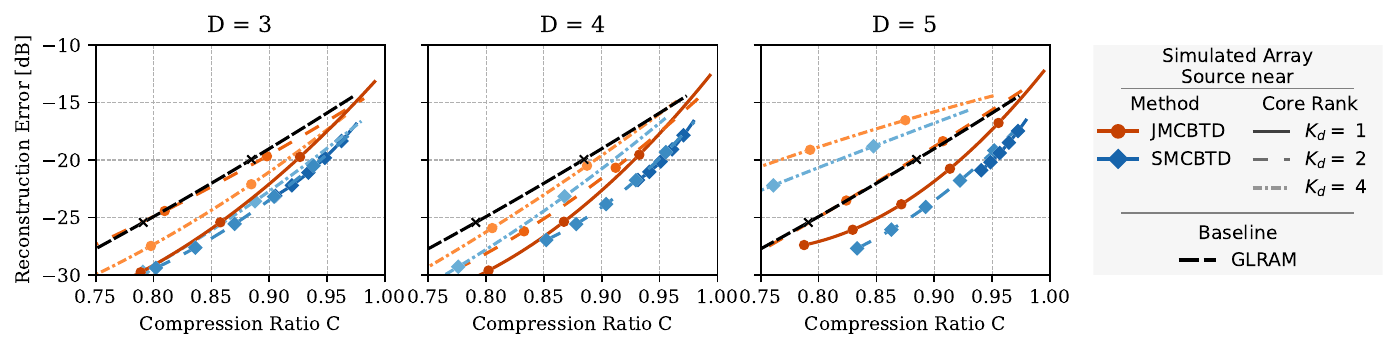}
    \caption[Compression rate vs. mean reconstruction error for different tensorization
    orders on simulated compact microphone array.]{
        Compression rate vs.\ mean reconstruction error for different tensorization orders
        \(D\) for a simulated compact microphone array at a near-field source position,
        see \autoref{fig:joint_comp:spatial_error}.  
        We compare the proposed \gls{jmcbtd} and \gls{sbtd} with the \gls{glram}-based method in
        \cite{jalmbyMultiChannelLowRankConvolution2024}.  
    }
    \label{fig:joint_comp:simulated-scenario-near}
\end{figure*}

\begin{figure*}
    \centering
    \includegraphics[width=\textwidth]{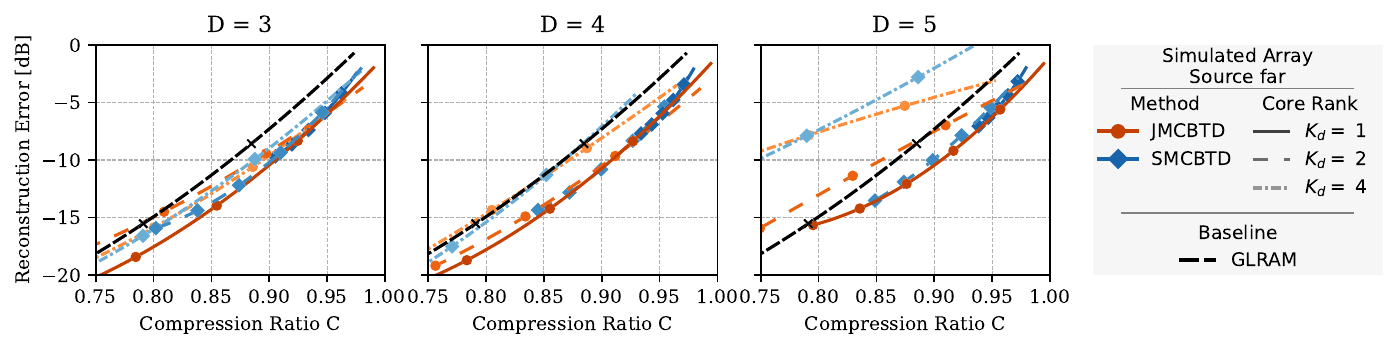}
    \caption[Compression rate vs. mean reconstruction error for different tensorization
    orders on simulated compact microphone array.]{
        Compression rate vs.\ mean reconstruction error for different tensorization orders
        \(D\) for a simulated compact microphone array at a far-field source position,
        see \autoref{fig:joint_comp:spatial_error}.  
        We compare the proposed \gls{jmcbtd} and \gls{sbtd} with the \gls{glram}-based method in
        \cite{jalmbyMultiChannelLowRankConvolution2024}.  
    }
    \label{fig:joint_comp:simulated-scenario-far}
\end{figure*}

\autoref{fig:joint_comp:simulated-scenario-near} and
\autoref{fig:joint_comp:simulated-scenario-far} show the mean reconstruction error versus
compression rate for different tensorization orders \(D \in \{3, 4, 5\}\) for a simulated
compact microphone array at a near- and far-field scenario.  
Additionally, the core rank is varied as \(K_d \in \{1, 2, 4\}\) for all modes \(d\).  
We can observe that at constant compression rate, increased core rank leads to worse
performance, with this effect being more pronounced for higher tensorization orders.  
This simulation provides numerical evidence for the low-rank \gls{cp} model of tensorized
\glspl{rir} that was introduced in
\autoref{sec:joint_comp:pole-zero-model-and-low-rank-tensor-representation}.  

\subsection{Measured Room Impulse Responses}
\label{sec:joint_comp:measured-room-impulse-responses}
The second set of experiments is conducted using measured \glspl{rir} from the
\gls{smard}, which contains \glspl{rir} measured in a small room using compact microphone
arrays and \gls{sriracha}, which contains \glspl{rir} measured in a shoebox room with
densely placed source positions over a grid covering large part of the room.  
Further, we use \gls{churchir}, to evaluate the performance in a highly reverberant
environment where \glspl{rir} exhibit a very dense modal structure
\cite{kuttruffRoomAcoustics2009}.  
We use the parameters listed in \autoref{tab:parameters-experiments} for the
proposed algorithm.  

\subsubsection{SMARD}
\label{sec:joint_comp:sim-smard}
The \gls{smard} dataset \cite{nielsenSingleMultichannelAudio2014} contains \glspl{rir}
measured for different setups of compact microphone arrays and loudspeakers in a small
room.  
For this, we select the subset of configurations using the directional loudspeaker and
group the \glspl{rir}.  
The grouping is done such that each group contains \glspl{rir} measured using the same
loudspeaker position but different microphone array setups.  
By the \gls{smard} naming convention, we denote these grouped configurations as 2X0X
grouping all configurations where X is any digit \(\{0,1,2\}\).  
This results in \(M=168\) channels.  
All \glspl{rir} are measured at a sampling frequency of \(f_s = 48\) kHz and truncated as
described at the beginning of \autoref{sec:joint_comp:numerical-simulations}.  
The average reverberation time is \(T_{60} \approx 0.15\)~s.  

\begin{figure*}
    \centering
    \includegraphics[width=\textwidth]{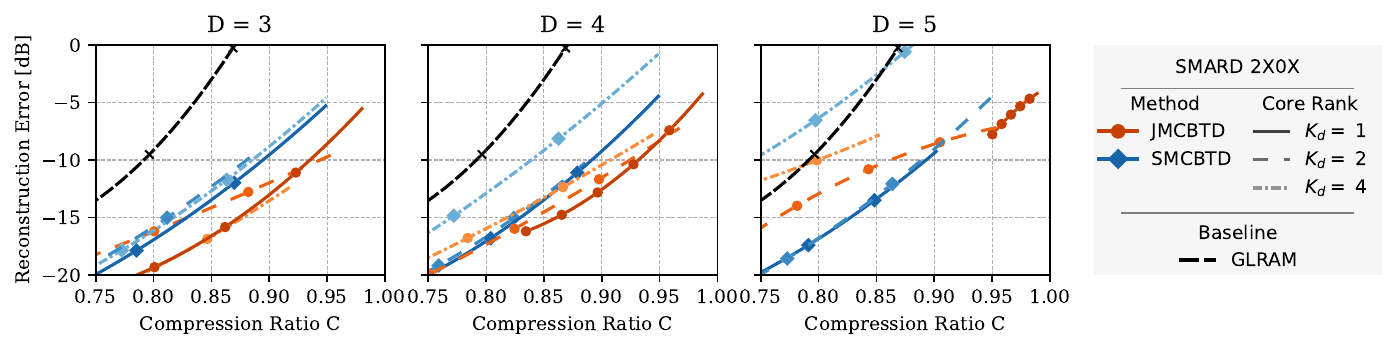}
    \caption[Compression rate vs. mean reconstruction error for different tensorization orders on \gls{smard}]{
        Compression rate vs.\ mean reconstruction error for different tensorization
        orders \(D\) on the set of impulse responses in \gls{smard}
        \cite{nielsenSingleMultichannelAudio2014} using all \glspl{rir} in the 2X0X
        configuration.  
        The performance gain of the proposed methods is substantial especially at \(K_d =
        1\), across all tensorization orders \(D=3,4,5\).  
        }
    \label{fig:joint_comp:smard-2x0x}
\end{figure*}
\autoref{fig:joint_comp:smard-2x0x} shows the mean reconstruction error versus compression ratios for
different tensorization orders \(D \in \{3, 4, 5\}\).  
The proposed method's performance gain is larger in this dataset compared to the simulated
scenario across all tensorization orders, especially at \(K_d = 1\) with a reconstruction
error gain of around 8 to 10~dB over the \gls{glram}-based method.  
Again, we can observe that at constant compression rate, increased core rank leads to
decreased performance, with this effect being more pronounced for higher tensorization
orders.  
It is worth noting that the \gls{jmcbtd} method's compression rate is computed after
spectral clustering which leads to a smaller range of compression rates visualized in the
plots.

\subsubsection{SRIRACHA}
\label{sec:joint_comp:sim-sriracha}
The \gls{sriracha} dataset \cite{pellingSRIRACHAShoeboxRoom2025} contains a large set of
measured \glspl{rir} in a shoebox room with densely placed source and receiver positions.  
We select two microphone positions, one at the center of the source position grid and one
on the edge of the grid, and use an equally spaced subset of \(M=256\) source positions
for each microphone position.  
All \glspl{rir} are provided at a sampling frequency of \(f_s = 36\)~kHz and are already
truncated to the average \(T_{60} \approx 0.5\)~s at 16,000 samples.  
The truncation process described at the beginning of \autoref{sec:joint_comp:numerical-simulations}
therefore may pad the \glspl{rir} with zeros to reach a more equidimensional tensor
shape.  

\begin{figure*}
    \centering
    \includegraphics[width=\textwidth]{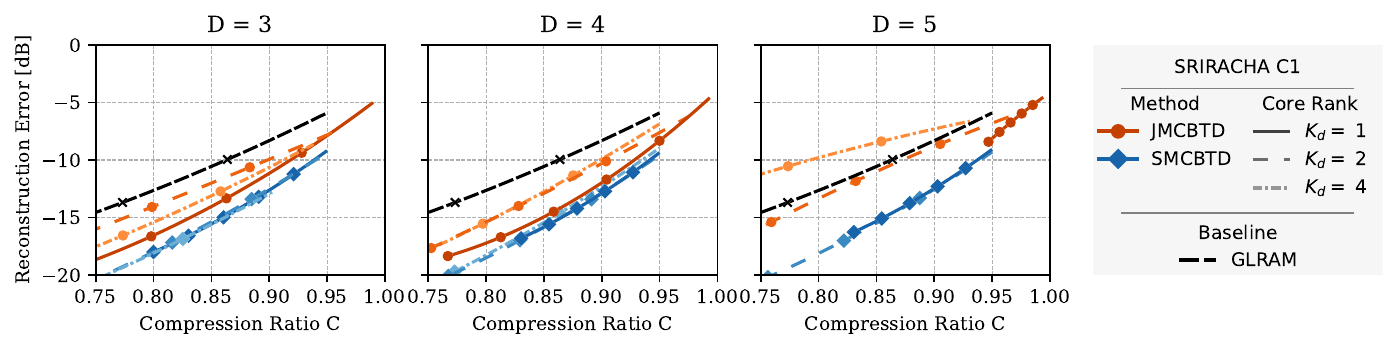}
    \caption[Compression rate vs. mean reconstruction error for different tensorization orders on SRIRACHA C1 dataset]{
        Compression rate vs.\ mean reconstruction error for different tensorization orders
        \(D\) on the SRIRACHA C1 dataset, spatially downsampled to \(M=256\) positions.  
        The performance gain of the proposed \gls{jmcbtd} and \gls{sbtd} methods is
        substantial especially at \(K_d = 1\), across all tensorization orders
        \(D=3,4,5\). Here \gls{sbtd} performs slightly better than \gls{jmcbtd}.
        }
    \label{fig:joint_comp:sriracha-c1}
\end{figure*}
\autoref{fig:joint_comp:sriracha-c1} shows the mean reconstruction error versus compression ratios for
different tensorization orders \(D \in \{3, 4, 5\}\).  
The proposed \gls{jmcbtd} and \gls{sbtd} methods outperform the \gls{glram}-based method across all
tensorization orders, especially at core rank \(K_d = 1\) with a reconstruction error gain
of around 5 to 8~dB.  
The effect of increased core rank leading to decreased performance at constant compression
rate is not as pronounced in this dataset compared to the previous ones.  
Also, the clustering effect in the shared factors found by the proposed \gls{jmcbtd} method
is less pronounced compared to the previous datasets.  
A potential reason for this is that this room, while being a shoebox, exhibits a denser
modal structure than the room used in \gls{smard}.  
It is worth noting that the \gls{jmcbtd} method's compression rate is computed after
spectral clustering which leads to a smaller range of compression rates visualized in the
plots.

\subsubsection{ChurchIR}
\label{sec:joint_comp:sim-churchir}
The \gls{churchir} dataset contains \glspl{rir} measured in a
church using higher-order spherical microphone arrays.  
We select two higher order arrays (Eigenmike) of 32 channels each and jointly compress
this set of \(M=64\) \glspl{rir}.  
Specifically, we select source position "SC" and receiver positions "E1" and "E2" as
described in \cite{giampiccolo2025churchir}.  
The room being a church, the reverberation time is very high with an average of \(RT_{30}
= 4.12\)~s according to \cite{giampiccolo2025churchir}.  
The authors of the dataset provide truncated \glspl{rir} to 7.0~s.
All \glspl{rir} are provided at a sampling frequency of \(f_s = 48\)~kHz, which we
downsample to \(f_s = 36\)~kHz.    
\begin{figure*}
    \centering
    \includegraphics[width=\textwidth]{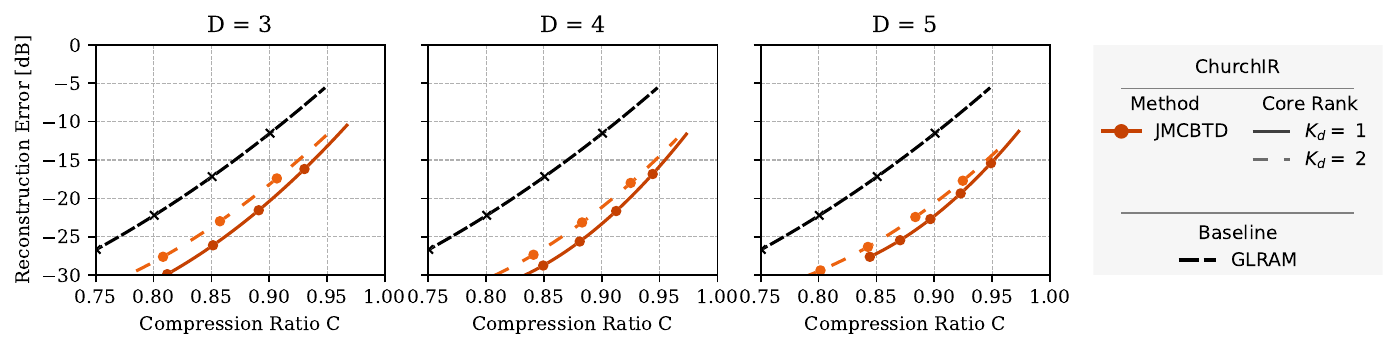}
    \caption[Compression rate vs. mean reconstruction error for different tensorization
    orders on the ChurchIR dataset]{
        Compression rate vs. mean reconstruction error for different tensorization orders
        \(D\) on a selection of \(M=64\) \glspl{rir} from the \gls{churchir} dataset.  
        The performance gain of the proposed \gls{jmcbtd} method is
        substantial especially at \(K_d = 1\), across all tensorization orders
        \(D=3,4,5\).  
        Here, the advantage of the \gls{jmcbtd} method over \gls{sbtd} is substantial as
        the temporary memory requirements of \gls{sbtd} make it infeasible to be applied
        to such long \glspl{rir}.  
        }
    \label{fig:joint_comp:churchir}
\end{figure*}
\autoref{fig:joint_comp:churchir} shows the mean reconstruction error versus compression ratios for
different tensorization orders \(D \in \{3, 4, 5\}\).  
The proposed \gls{jmcbtd} method outperforms the \gls{glram}-based method
across all tensorization orders, again, especially at core rank \(K_d = 1\) with a
reconstruction error gain of around 10~dB.  
The proposed \gls{sbtd} method becomes infeasible to apply in this scenario due to the
temporary memory requirements growing beyond available working memory, see
\autoref{sec:joint_comp:computational-complexity}.  
The effectiveness of the methods in this highly reverberant scenario is more pronounced
than in the previous datasets.  
So despite the denser modal structure of the \glspl{rir} in this dataset, its relationship
to the much larger reverberation time and therefore higher sample counts, still allows for
a low-rank tensor representation after tensorization.  
Further, here, increasing the number of modes \(D\) seems to have a more pronounced
effect on the performance compared to the previous datasets, where a gain of around 2~dB
can be observed when going from \(D=3\) to \(D=4\) at core rank \(K_d = 1\).  

\subsection{Tail Reconstruction}
\label{sec:joint_comp:tail_reconstruction}
To complement the standard reconstruction error, which is dominated by the high-energy
direct sound and early reflections, we additionally evaluate the late reverberant part of
the impulse response using a log-envelope error measure. Let \(n_{\text{tail}}\) denote
the starting sample of the tail, corresponding to the mixing time \(t_{\text{mix}} \approx
\sqrt{V}\) in ms \cite{defrance2008measuring}, used as a reasonable approximation of the
transition from early reflections to late reverberation.  
For an impulse response \(h[n]\), we first compute a smoothed local-energy envelope
\begin{equation}
    e_h[n] = (h^2[n]) \ast w[n]
\end{equation}
where \(w[n]\) is a short smoothing window.  
The corresponding tail-start normalized envelope is
\begin{equation}
    \tilde{E}_h[n] = 10 \, \log_{10} (e_h[n]) - E_h[n_{\text{tail}}], \quad \forall \, n.
\end{equation}
This is computed analogously for the reconstructed \gls{rir} \(\hat{h}[n]\), and
we then define the \gls{tee} as 
\begin{equation}
    \text{TEE} = \frac{1}{L - n_{\text{tail}}} \sum_{n=n_{\text{tail}}}^{L-1} | \tilde{E}_h[n] - \tilde{E}_{\hat{h}}[n] |.
\end{equation}
This metric is expressed in dB and specifically quantifies how well the local decay
profile of the late reverberant tail is preserved, which corresponds better to diffuse
late reverberation than sample-wise misalignment.  

\begin{figure}
    \centering
    \includegraphics[width=\columnwidth]{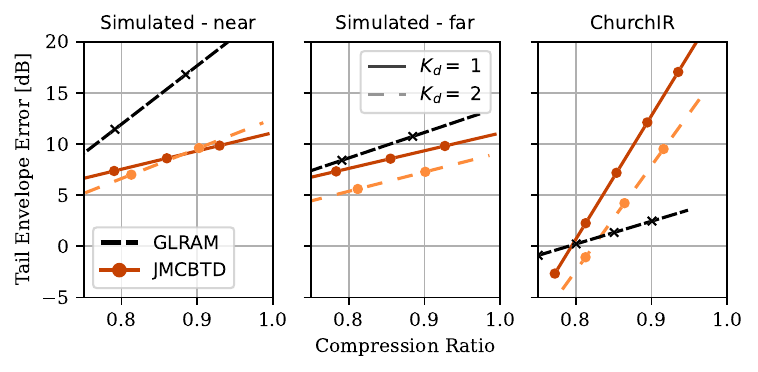}
    \caption[Compression rate vs. tail envelope error]{
        Compression rate vs. \gls{tee} on the simulated and \gls{churchir} datasets
        representing short and long \glspl{rir}, taking into account the
        tail from the mixing time at 14~ms and 171~ms, respectively. Lower values represent
        better tail reconstruction.  
        }
    \label{fig:joint_comp:tee}
\end{figure}

\autoref{fig:joint_comp:tee} shows
the \gls{tee} for the simulated dataset and \gls{churchir} plotted over compression ratio. 
We use a rectangular smoothing window of 10~ms length and use the mixing times as listed
in \autoref{tab:parameters-experiments}.  
We limit the investigation to two datasets and a comparison between the baseline and
\gls{jmcbtd} with core sizes \(K_d \in {1,2}\) for the simulated data and \gls{churchir}
to represent short and long \glspl{rir}, respectively.  
Compared to the reconstruction error, where scalar cores show the best performance, here the
trend is that larger cores enable better tail reconstruction.  
Further, for \gls{churchir}, the baseline \gls{glram} performs better at higher
compression ratios, while the proposed method does so for lower compression ratios.  

\subsection{Floating-Point Precision}
\label{sec:joint_comp:float}
\begin{figure}
    \centering
    \includegraphics[width=\columnwidth]{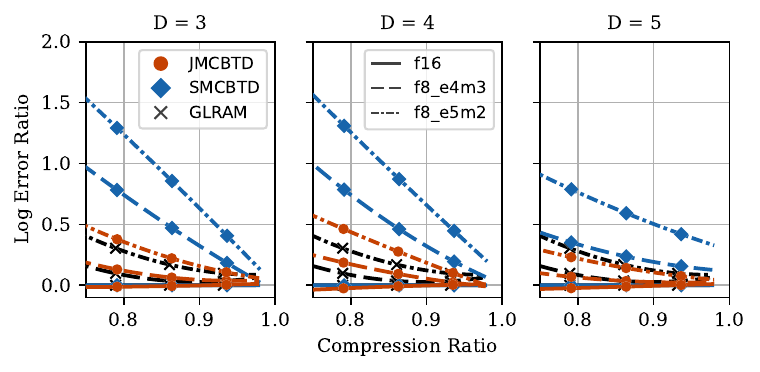}
    \caption[Sensitivity to floating-point precision]{
        The \gls{ler} \eqref{eq:joint_comp:error_ratio} for tensorization orders \(D
        \in \{3,4,5\}\) over compression ratio \(C\) for the algorithms \gls{jmcbtd},
        \gls{sbtd} and the baseline \gls{glram}. 
        }
    \label{fig:joint_comp:error_ratio}
\end{figure}

We perform a simple check to test the decomposition's sensitivity to decreasing
floating-point precision. For this, we convert the tensor decompositions for \gls{jmcbtd},
\gls{sbtd}, and \gls{glram} from the original 32-bit floating-point representation into
16- and 8-bit formats and compute the reconstruction error
\eqref{eq:joint_comp:normalized-misalignment} w.r.t the original \glspl{rir}.  
We compute the \gls{ler}
\begin{equation}
    \text{LER}_{\{16,8\}} = \log_2 \frac{\overline{\text{NM}}_{\{16,8\}}}{\overline{\text{NM}}_{32}}
    \label{eq:joint_comp:error_ratio}
\end{equation}
for the following PyTorch \cite{anselPyTorch2Faster2024} floating-point types:
\texttt{float16}, \texttt{float8\_e4m3} (4-bit exponent; 3-bit mantissa), and
\texttt{float8\_e5m2} (5-bit exponent; 2-bit mantissa), with the \texttt{float32} type as
the baseline.  
This ratio, computed for decompositions at various compression ratios, shows how large the
error induced by the decreased floating point precision is compared to the original
high-precision decomposition.  
The logarithm symmetrizes the values w.r.t. the unit ratio.  
A \gls{ler} of 0 indicates no reconstruction loss due to reduced floating-point precision,
while a value of 1 indicates the reconstruction error induced by precision and
decomposition are equally large.  
In \autoref{fig:joint_comp:error_ratio}, we show the \glspl{ler} for the proposed and
baseline methods for the simulated \gls{rir} scenario.  
We can observe the \gls{ler} decreases with higher compression ratios.  
This implies that the error induced by the decreased precision becomes less significant at
higher compression ratios compared to the error induced by the compressed representation
itself. 
Generally, 16-bit precision does not introduce additional error for all three methods,
while for 8-bit precision, for all methods, \texttt{float8\_e4m3} shows larger \gls{ler}
than \texttt{float8\_e5m2}.  
\gls{sbtd} is affected more significantly than \gls{jmcbtd} and the baseline \gls{glram}.  

\section{Discussion}
\label{sec:joint_comp:discussion}
One important aspect of the proposed methods is that the compression not just reduces the
number of parameters needed to represent the \glspl{rir}, but also allows for efficient
downstream applications such as convolution with anechoic signals.  
The efficiency of such convolution operations is described in detail in
\cite{jalmbyMultiChannelLowRankConvolution2024} for the \gls{cp} and \gls{glram} methods.  
Since \gls{btd} generalizes \gls{cp}, similar computational savings can be achieved.  

The floating-point precision experiment suggests that the proposed decompositions can be
stored in 16-bit precision with negligible additional reconstruction error, making
reduced-precision storage a promising complement to tensor-based compression.  
More aggressive 8-bit quantization appears more sensitive to the chosen format and is left for
future optimization.


The numerical experiments in \autoref{sec:joint_comp:numerical-simulations} show that the proposed
\gls{jmcbtd} method outperforms existing methods for joint compression of multichannel
\glspl{rir} across a range of compression rates and tensorization orders.  
While the performance gain is significant from \(D=2\) to \(D=3\) in all experiments, the
gain from \(D>3\) is marginal to none in most scenarios.  
This indicates that a tensorization order of \(D=3\) is sufficient to capture the
multilinear structure of \glspl{rir} for compression purposes.  
Similar improvements were observed for the \gls{sbtd} method.  
The exception is the \gls{churchir} dataset, where increasing the tensorization order from
\(D=3\) to \(D=4\) shows a more pronounced performance gain.  
This leads to the hypothesis that the method's effectiveness increases with increasing
\gls{rir} lengths.  

The numerical results underline the theoretical correspondence between \gls{pz} model and
low-rank tensor representations, and point to the \gls{cp} as the most efficient
representation in terms of reconstruction error.  
Nonetheless, the more general \gls{btd} shows better tail reconstruction in some cases.  
This, combined with the potential to further investigate multilinear low-rank structures,
i.e., cores that are larger than scalar, for data-driven rank estimation instead of rank
being a fixed hyperparameter keep the \gls{btd} a valuable representation.  

In implementation, there is a trade-off between computation time and temporary memory
requirement to make: intermediate residuals can either be kept in memory or
recomputed on-demand.  
This number grows with both increasing tensorization order \(D\) and number of blocks
\(R\).  
In all experiments, the range of compression ratios covered shrinks with increasing
tensorization order \(D\) due to this effect as the feasibility of the computations is
limited by available memory (and time).  
Here, the difference between \gls{jmcbtd} and \gls{sbtd} is that the former requires
smaller intermediate residuals due to the smaller working sets in each block update.  
This makes the \gls{jmcbtd} method more scalable for higher channel counts \(M\),
\gls{rir} lengths \(L\), and number of blocks \(R\).  
The fact of increasing computation time and temporary memory requirement with increasing
tensorization order \(D\), combined with the marginal performance gain in terms of
reconstruction error observed in the experiments, suggests that \(D=3\) is a good choice
in practice.  


Future extensions of the proposed methods are a single step clustering approach for the
\gls{jmcbtd} avoiding the two-step approach, and data-driven detection of the multilinear
rank for both \gls{jmcbtd} and \gls{sbtd}, which was considered a fixed hyperparameter
for this work.  
Although scalar cores yield the best compression performance in our experiments, they
may still provide valuable insight into multichannel \gls{rir} tensor modelling.  
Further, while the clustering approach used in the \gls{jmcbtd} method shows good
performance gains, it is currently not possible to predict the clustering
parameters for a certain target compression rate and reconstruction error, therefore
further research is needed to address this challenge.

\section{Conclusion}
\label{sec:joint_comp:conclusion}
In this paper, we first introduced \gls{btd} for the compression of acoustic \glspl{rir},
generalizing previous work on \gls{cp} decomposition and \gls{glram}-based methods.  
We showed how the \gls{pz} model leads to a low-rank tensor when reshaping a single
\gls{rir} into a higher-order tensor, which confirms empirical findings of existing work,
bridging a gap in the literature. We then proposed two methods for joint compression of
multichannel \glspl{rir} using the \gls{btd}.  
The first method \gls{sbtd} compresses a stacked tensor of \glspl{rir}, while the second
\gls{jmcbtd} method considers shared factors across \glspl{rir}.  
We proposed a two-step approach using spectral clustering to find optimal clusters of
\glspl{rir} that share factors.  
Numerical experiments using simulated and measured \glspl{rir} showed improved
performance of the proposed methods over the baseline method.